\documentclass[journal]{IEEEtran}

\usepackage{times}
\usepackage{epsfig}
\usepackage{amssymb}
\usepackage{subfigure}
\usepackage{bbding}
\usepackage{xcolor}
\usepackage{threeparttable}
\usepackage{booktabs}
\usepackage{algorithm}
\usepackage{algorithmic}

\newcommand{\eg}{\textit{e}.\textit{g}.}
\newcommand{\ie}{\textit{i}.\textit{e}.}

\usepackage{url}

\usepackage{listings}
\usepackage{warpcol}
\usepackage{makecell}
\usepackage{amsthm,amsmath,amssymb}
\usepackage{mathrsfs}
\newcommand{\etal}{\textit{et al}.}

\usepackage{graphicx, color, multirow, amsmath}

\RequirePackage{latexsym,amsmath,amssymb}

\ifCLASSINFOpdf
\else
\fi

\begin{document}

\title{Multitask Auxiliary Network for Perceptual Quality Assessment of Non-Uniformly Distorted Omnidirectional Images}
\author{Jiebin Yan,
    Jiale Rao,
    Junjie Chen,
    Ziwen Tan,
    Weide Liu,
    Yuming Fang,~\IEEEmembership{Senior~Member,~IEEE}

\thanks{This work was supported in part by the National Key Research and Development Program of China under Grant 2023YFE0210700, in part by the National Natural Science Foundation of China under Grants 62461028, 62441203, 62311530101 and 62132006, in part by the Natural Science Foundation of Jiangxi Province of China under Grants 20243BCE51139, 20242BAB21006 and 20232BAB202001, and in part by the project funded by China Postdoctoral Science Foundation under Grant 2024T170364. (Corresponding author: Yuming Fang).}
    
\thanks{Jiebin Yan, Jiale Rao, Junjie Chen, Ziwen Tan, and Yuming Fang are with the School of Computing and Artificial Intelligence, Jiangxi University of Finance and Economics, and also with Jiangxi Provincial Key Laboratory of Multimedia Intelligent Processing, Nanchang 330032, China (e-mail: yanjiebin@jxufe.edu.cn, jialerao@foxmail.com, chenjunjie@jxufe.edu.cn, ziwentan@foxmail.com, fa0001ng@e.ntu.edu.sg).}
\thanks{W. Liu is with Harvard Medical School, Harvard University, USA (e-mail: weide001@e.ntu.edu.sg).}
}


\maketitle

\begin{abstract}

Omnidirectional image quality assessment (OIQA) has been widely investigated in the past few years and achieved much success. However, most of existing studies are dedicated to solve the uniform distortion problem in OIQA, which has a natural gap with the non-uniform distortion problem, and their ability in capturing non-uniform distortion is far from satisfactory. To narrow this gap, in this paper, we propose a multitask auxiliary network for non-uniformly distorted omnidirectional images, where the parameters are optimized by jointly training the main task and other auxiliary tasks. The proposed network mainly consists of three parts: a backbone for extracting multiscale features from the viewport sequence, a multitask feature selection module for dynamically allocating specific features to different tasks, and auxiliary sub-networks for guiding the proposed model to capture local distortion and global quality change. Extensive experiments conducted on two large-scale OIQA databases demonstrate that the proposed model outperforms other state-of-the-art OIQA metrics, and these auxiliary sub-networks contribute to improve the performance of the proposed model. The source code is available at \url{https://github.com/RJL2000/MTAOIQA}.
\end{abstract}


\IEEEpeerreviewmaketitle

\section{Introduction}
\label{sec:intr}

\IEEEPARstart{W}{ith} the vigorous development of artificial intelligence technology and hardware devices, virtual reality (VR) applications have become widespread in our daily life, such as tourism, education, and entertainment. Omnidirectional image (OI), as the core medium of VR technology, is crucial for users to experience immersive visual content~\cite{IQASurvey}. Generally, users can freely explore immersive content by feat of head-mounted displays (HMDs). However, during the process of generation, transmission, displaying, postprocessing, and \emph{etc}, inevitable distortions may occur to varying degrees~\cite{End2EndQoE}. Viewing degraded OIs through HMDs may cause users physical and psychological discomfort~\cite{lin2002effects}. Therefore, accurately measuring the quality of OIs is essential for guiding VR system improvement and algorithm optimization~\cite{ding2021comparison,fang2021superpixel,le2021perceptually}.

\begin{figure}[t]
\centering
\includegraphics[width=1\linewidth]{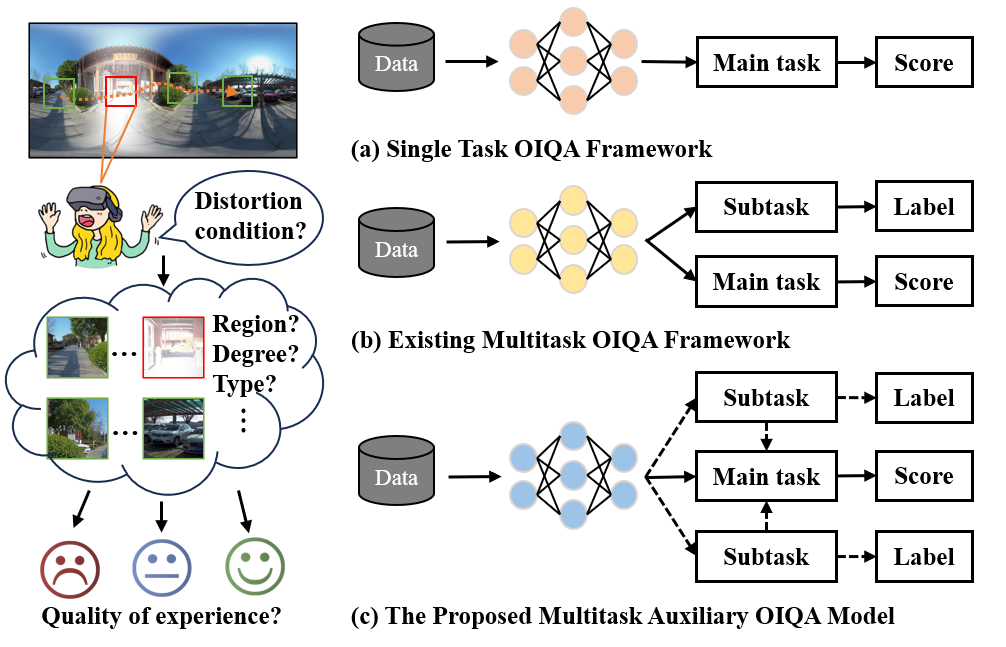}
\caption{An intuitive comparison between the proposed multitask auxiliary OIQA model, single task OIQA framework, and existing multitask OIQA framework.}
\label{fig:intro}
\end{figure}

In general, omnidirectional image quality assessment (OIQA) includes subjective OIQA and objective OIQA, where the former refers to conducting psychophysical experiments, from which we can analyze the effect of those influencing factors on image quality qualitatively. However, the subjective procedures are always labor-intensive and expensive, and hard to be embedded in real application. On the contrary, objective OIQA methods can automatically quantify image degradation and are easily adapted to nearly all parts in the whole multimedia system, and therefore are more desired. Depending on whether reference image is accessible, objective OIQA models can be divided into full-reference (FR-OIQA) and no-reference/blind (BOIQA) methods. The former require reference information, while the latter can predict the quality of OIs without reference information. Most FR-OIQA methods~\cite{s-psnr,cpp-psnr,ws-psnr,s-ssim,ws-ssim} are based on mature 2D-IQA methods such as PSNR and SSIM~\cite{ssim}, and are improved by incorporating the spherical characteristics of OIs. However, their performance is still far from satisfactory due to the diverse users' viewing behaviors and complicated brain activity. Recently, many deep learning OIQA methods~\cite{MC360IQA,Sui,fang2022perceptual,wu2023assessor360} have been successively proposed, leading to significant improvement, where these methods simulate user viewing behavior explicitly or implicitly and can learn the complicated relationship between image degradation and visual quality by deep learning.

Although objective OIQA models have shown continuous progress, some shortcomings remain. First, their success in measuring the quality of uniformly distorted OIs (\ie, the entire OI is affected by the ``same amount'' of distortion) cannot be transferred to OIs with locally distributed non-uniform distortion~\cite{fang2022perceptual} (\ie, some regions of the OI are affected by a ``different amount'' of distortion compared to other regions), since there exists a natural gap between uniform distortion and non-uniform distortion. Second, they neglect the role of auxiliary tasks, which are conductive to capture local quality degradation and global quality change. As shown in Fig.~\ref{fig:intro}, most OIQA methods are trained as in a single-task framework, disregarding the important information from auxiliary tasks. Some multitask OIQA methods only utilize related tasks by joint learning, overlooking the interaction between the auxiliary tasks and the main task. According to the study on cognitive psychology~\cite{kirkpatrick2017overcoming}, the human brain exhibits characteristics of multitask auxiliary learning. In addition, users can easily identify distortion range (\eg, two regions are distorted), distortion type (\eg, noticeable brightness changes and noise distribution in OIs) as well as distortion degree. Inspired by this, in this paper, we propose a multitask auxiliary learning model for non-uniformly distorted OIs. To this end, we meticulously design a multitask feature selection module, a series of auxiliary sub-networks, and a multitask auxiliary fusion module, where the former adaptively selects specific features as input for different tasks, the middle guide the proposed model to capture local distortion and global quality change, and the latter integrates information from auxiliary tasks with the features of the main task.

In summary, our contributions are three fold.
\begin{itemize}

\item To better learn generalized features for capturing locally distributed non-uniform distortion, we propose a multitask BOIQA framework by introducing three quality-relevant auxiliary tasks, including distortion range prediction, distortion type prediction, and distortion degree prediction, which are used to guide the proposed model to pay partial attention to global quality change and local distortion.

\item To dynamically allocate specific features to different tasks, we design a novel multitask feature selection module, where the fused multi-scale features are adopted by each task with learnable probabilities. Besides, we propose a series of attention models to fit different tasks.

\item To demonstrate the effectiveness of the proposed model and these specifically designed modules, we conduct comprehensive experiments on two large-scale OIQA databases, where we verify that the proposed BOIQA model outperforms the state-of-the-art methods and the newly proposed modules contribute to performance improvement.

\end{itemize}

\begin{figure*}[t]
\centering
  \includegraphics[width=1\linewidth]{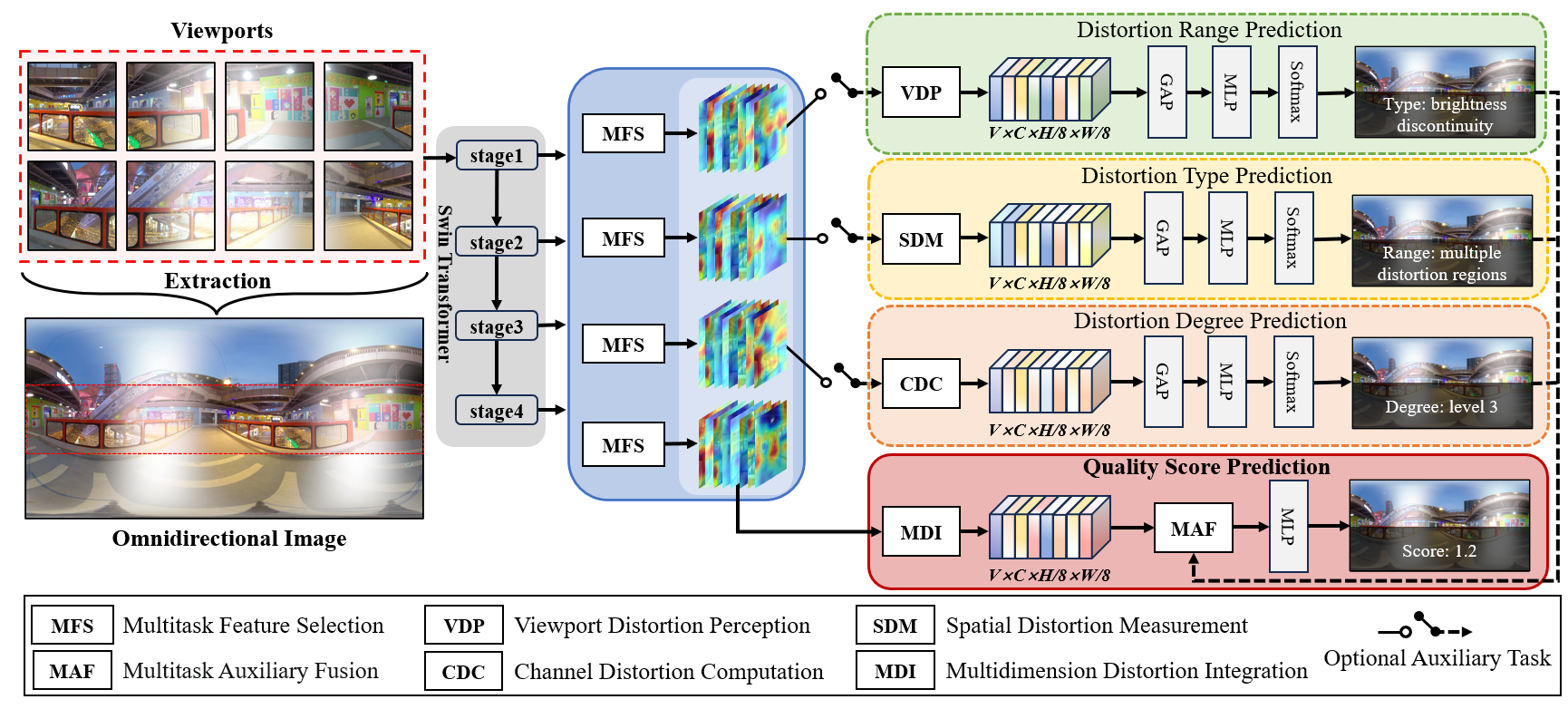}
\caption{The framework of the proposed model. It consists of three optional auxiliary tasks and a main task, where three auxiliary tasks are distortion range prediction, distortion type classification, and distortion degree classification, respectively, and the main task is quality score prediction. To connect these tasks seamlessly, a MFS module is used to adaptively select specific features for different tasks, and a MAF module is used to combine the features from different tasks for global quality prediction.}
\label{fig:framework} 
\end{figure*}

\section{Related Work}
\label{sec:related}

In this section, we review the blind IQA (BIQA) and OIQA models, respectively.

\subsection{BIQA Models}
\label{subsec:2d-iqa}

Generally, traditional BIQA models mainly include two steps: feature extraction and quality regression, where feature extraction is particularly critical~\cite{fang2020blind},~\cite{xian2023perceptual}. Mittal~\etal~\cite{mittal2012making} proposed an unsupervised learning-based model called natural image quality evaluator (NIQE), where quality-aware features are derived from the natural scene statistics (NSS), and the distance between the multivariate Gaussian model statistics of the distorted image and high-quality images is used to represent the quality score. Rather than fitting feature distribution, Fang~\etal~\cite{fang2017no} and Yan~\etal~\cite{yan2020no} proposed using the histogram to represent the local to global change of screen content images and 3D synthesized images, respectively.

Different from those traditional models, deep learning-based models~\cite{zhou2024hdiqa},~\cite{shen2024graph},~\cite{zhou2022end} bridge the gap between these two separate steps, \ie, feature extraction and quality regression, and can be obtained by end-to-end learning. Kang~\etal~\cite{kang2014convolutional} presented an early study by introducing the convolutional neural network (CNN) in IQA. Kim~\etal\cite{kim2017deep} tested these image classification networks in IQA and found that their success in image classification can be easily transferred to IQA. Later, many researchers tried to improve the performance of BIQA models from different points of view. Su~\etal~\cite{su2020blindly} proposed an adaptive hyper network architecture that considers image content comprehension during perceptual quality prediction. Yang~\etal~\cite{yang2022maniqa} proposed a multidimensional attention model, which designs a transposed attention block and a scale swin transformer block for multidimensional interaction in channel and spatial domains. Pan~\etal~\cite{pan2022vcrnet} proposed a visual compensation restoration model, which consists of a visual restoration network and a quality assessment network. Yi~\etal~\cite{yi2023towards} employed a combination of style-specific and generic aesthetic characteristics to evaluate image artistic quality. To solve the uncertainty problem across different databases, Zhang~\etal~\cite{zhang2021uncertainty} leveraged continuous ranking information from mean opinion scores and the difference between human opinion scores to train a unified learning-to-rank IQA model. Except for those BIQA models with only one task, many studies also attempted to introduce more information to guide model training. Kang~\etal~\cite{kang2015simultaneous} designed a compact CNN model to simultaneously learn distortion type and image quality information. Considering the relevance between these two tasks, Ma~\etal~\cite{ma2017end} proposed a multitask network in which the probability of distortion type is used to assist quality prediction. Madhusudana~\etal~\cite{madhusudana2022image} proposed an unsupervised training method that treats classification as an auxiliary task and uses contrastive learning to train a quality-aware network. Zhang~\etal~\cite{zhang2023blind} proposed a multitask scheme which considers the correspondence between vision and language.

\subsection{OIQA Models}

A na\"{\i}ve way to evaluate the quality of omnidirectional images is adapting 2D-IQA models, \eg, PSNR and SSIM~\cite{ssim}, to this task by considering the sphere characteristic of omnidirectional images, such as S-PSNR~\cite{s-psnr}, CPP-PSNR~\cite{cpp-psnr}, WS-PSNR~\cite{ws-ssim}, S-SSIM~\cite{s-ssim}, and WS-SSIM~\cite{ws-ssim}. However, these models are far from being satisfactory, since they overlook the fact that only a limited field of view can be seen at a certain moment. Afterward, more and more advanced OIQA models have been proposed, which can be roughly classified into two categories, \ie, patch-based models, and viewport-aware models. The former type of models always accept the patches generated by varied projection methods, \eg, equirectangular projection (ERP), as input. Kim~\etal~\cite{kim2019deep} proposed to use visual features of each patch and its position information in the format of ERP to predict patch weight and quality score, and then fuse the quality scores of all patches to get the overall quality score. Jiang~\etal~\cite{Jiang2021tip} extracted hand-crafted features in the format of cube map to represent omnidirectional image quality. Zhou~\etal~\cite{zhou2023perception} proposed to extract quality-aware features from both ERP and cube map projection formats, where the ERP image is used to extract just-noticeable difference features using a saliency method and the cube map is used to extract multiscale details, and these features are then used to predict image quality together. Li~\etal~\cite{li2023mfan} proposed a multi-projection fusion mechanism for OIQA which projects an OI into cubemap (CMP), pyramid projection (PYM), and ERP formats, and the extracted features from these three formats are used to predict final quality score using a polynomial regression. To avoid the contradiction between the geometry deformation inborn from ERP images and the computational cost on viewport generation, Yan~\etal~\cite{yan2024viewport} proposed a viewport-independent and deformation-unaware OIQA model, where an equirectangular modulated deformable convolution is designed to solve the deformation problem.

Different from the patch-based models, these viewport-aware models aim to incorporate users' observation process in quality prediction. Sun~\etal~\cite{MC360IQA} designed a multichannel network for OIQA, which accepts six viewports as input and uses the fused features of generated viewports to predict overall quality. Zhou~\etal~\cite{zhou2021omnidirectional} introduced an auxiliary task, \ie, distortion classification, to assist quality feature learning for viewport images. In fact, these two studies~\cite{MC360IQA, zhou2021omnidirectional} consider all regions equally, while ignoring users' viewing bias, \ie, pay more attention to salient regions or those near the equator. To address this problem, Xu~\etal~\cite{VGCN} designed a two-branch OIQA model, whose local branch accepts salient regions as input and uses a graph convolutional network to capture viewport-wise dependency. Fang~\etal~\cite{fang2022perceptual} proposed an OIQA model to measure the non-uniform distortion, which incorporates viewing condition (which is quantified by coordinates and exploration times) in feature representation of viewports. Liu~\etal~\cite{liu2024perceptual} further extended this work~\cite{fang2022perceptual} by mainly replacing the feature extraction module with a more powerful backbone and a transform encoder. Wu~\etal~\cite{wu2023assessor360} proposed to generate multiple pseudo viewport sequences using a recursive probability sampling method and designed a multiscale feature aggregation module and a temporal modeling module to extract quality-aware features of viewports and their temporal correlation. Different from other studies, Yang~\etal~\cite{yang2022tvformer} formulated viewing trajectory prediction and quality estimation in a unified framework.

\section{The Proposed OIQA Model}
\label{sec:abff}

In this section, we propose a MultiTask Auxiliary OIQA model named MTAOIQA to better address the issue of non-uniform distortion by obtaining relevant semantic features from other optional auxiliary tasks and fusing them with the quality score prediction task, and its architecture is shown in Fig.~\ref{fig:framework}. Specifically, MTAOIQA consists mainly of a backbone (\ie, swin transformer~\cite{liu2021swin}) for extracting multi-scale features of viewports, a multitask feature selection module for adaptively allocating specific features to different auxiliary tasks, a multitask auxiliary fusion module for deeply integrating features from different tasks, and a quality score prediction module. We then introduce these modules in detail.

\subsection{Viewport Feature Extraction}

Here, the priority is to extract viewports as the input. Considering the fact that users tend to prioritize viewing the low-latitude regions of OIs, we adopt an equatorial sampling method to simulate the user's viewing behavior. By doing this, we can obtain the viewports that cover the low-latitude regions. The specific procedure can be referred to~\cite{JVETG1003}. Given an OI denoted by $\mathbf{I}$, we can sample the $V$ viewports $\mathbf{VS}=\{\mathbf{I},\left\{\mathbf{V}_i,(\phi_i)\right\}_{i=1}^V\}$, where $\mathbf{V}_i$ denotes the $i$-th viewport image, $\phi_i$ is the coordinate of the $i$-th viewport image. Considering that the swin transformer exhibits exceptional feature learning capabilities and demonstrates strong generalization performance across various visual tasks, we use it as the network backbone $B$ for staged feature extraction of $\mathbf{VS}$, whose output consists of four representations: $\{\mathbf{F}_1 \in \mathbb{R}^{ V \times C1 \times \frac{H}{8} \times \frac{W}{8}}, \mathbf{F}_2 \in \mathbb{R}^{ V \times  C2 \times \frac{H}{16} \times \frac{W}{16} }, \mathbf{F}_3 \in \mathbb{R}^{ V \times  C3 \times \frac{H}{32} \times \frac{W}{32}}, \mathbf{F}_4 \in \mathbb{R}^{ V \times C4 \times \frac{H}{32} \times \frac{W}{32}}\}$, which can be formulated as
\begin{eqnarray}
\centering 
\mathbf{F}_s = \left\{(B(\mathbf{VS};\Theta_{B}))\right\}_{s=1}^4,
\label{eq:feature}
\end{eqnarray}
where $\Theta_{B}$ is the parameters of the backbone; $s$ denotes the index of four stages. 

\begin{figure}[t]
\centering
\includegraphics[scale=0.28]{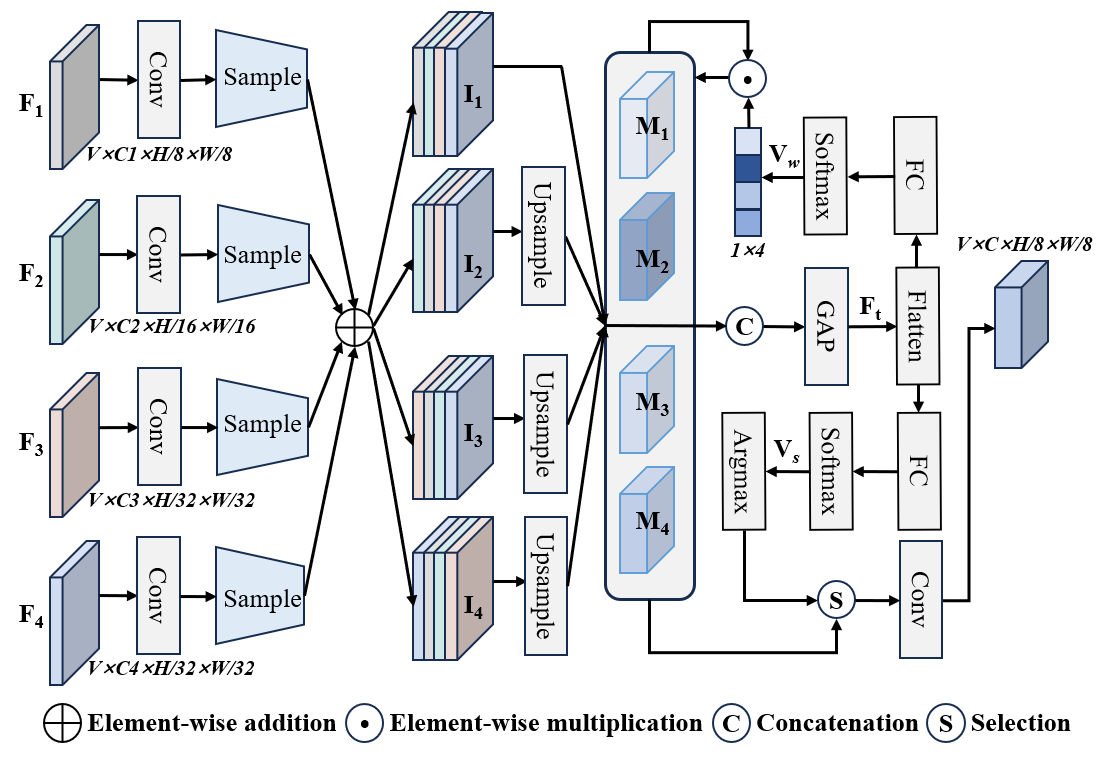}
\caption{The multitask feature selection module.}
\label{fig:mfs}
\end{figure}

\subsection{Multitask Feature Selection}

The human visual system can perceive features at multiple levels, enabling accurate identification and understanding of visual information in images. In many studies~\cite{lan2023multilevel, shen2022end, zhou2022attentional}, local and global information is obtained by utilizing multi-level feature extraction and fusion to improve the prediction of images. This inspires us to fuse multi-scale features and adaptively select features suitable for different tasks. To this end, we design a multitask feature selection (MFS) module, which is shown in Fig.~\ref{fig:mfs}. Specifically, $\{\mathbf{F}_s\}_{s=1}^4$ are adopted as input, and four $1\times1$ convolutions $\mathrm{W}_{\Bar{F}_1}$, $\mathrm{W}_{\Bar{F}_2}$, $\mathrm{W}_{\Bar{F}_3}$, and $\mathrm{W}_{\Bar{F}_4}$ are used to squeeze and unify channel numbers, then $\{\mathbf{\Bar{F}}_1 \in \mathbb{R}^{ V \times C \times \frac{H}{8} \times \frac{W}{8} }, \mathbf{\Bar{F}}_2 \in \mathbb{R}^{ V \times  C \times \frac{H}{16} \times \frac{W}{16}  }, \mathbf{\Bar{F}}_3 \in \mathbb{R}^{ V \times  C \times \frac{H}{32} \times \frac{W}{32}}, \mathbf{\Bar{F}}_4 \in \mathbb{R}^{ V \times C \times \frac{H}{32} \times \frac{W}{32}}\}$ can be obtained by
\begin{eqnarray}
\begin{split}
\centering 
&\mathbf{\Bar{F}}_1=\mathrm{W}_{\Bar{F}_1}(\mathbf{F}_1),\mathbf{\Bar{F}}_2=\mathrm{W}_{\Bar{F}_2}(\mathbf{F}_2),\\
&\mathbf{\Bar{F}}_3=\mathrm{W}_{\Bar{F}_3}(\mathbf{F}_3),\mathbf{\Bar{F}}_4=\mathrm{W}_{\Bar{F}_4}(\mathbf{F}_4),
\end{split}
\label{eq:feature}
\end{eqnarray}
where $C$ represents the number of channels after unification. Then, a cross-resolution fusion operation is designed to obtain the corresponding interaction feature $\mathbf{I}_m$, which can be expressed as 
\begin{eqnarray}
\centering
\left \{ \mathbf{I}_m\right \}_{m=1}^4 =\left \{\sum_{n=1}^{4} \mathrm{S}_{m,n}\left (\mathbf{\Bar{F}}_m, \mathbf{\Bar{F}}_n  \right ) \right \}_{m=1}^4, m \neq n
\label{eq:feature}
\end{eqnarray}
where $\mathrm{S}_{m,n}(\mathbf{\Bar{F}}_m,\mathbf{\Bar{F}}_n)$ denotes an operation sequence, including reshape $\mathbf{\Bar{F}}_n$ to the size of $\mathbf{\Bar{F}}_m$ and concatenate these resultant feature maps with $\mathbf{\Bar{F}}_m$; $\mathbf{I}_m$ represents the corresponding output.

To maintain the integrity of the information of $\{\mathbf{I}_m\}_{m=1}^4$, we perform upsampling operations on $\{\mathbf{I}_m\}_{m=2}^4$ to be of the same size as $\mathbf{I}_1$, and can obtain multilevel features $\{\mathbf{M}_m\}_{m=1}^4$. Then, a concatenation operation is applied, which is followed by a $1 \times 1 $ convolution in the channel dimension and a global average pooling (GAP) in the spatial dimension. The whole procedure can be formulated by 
\begin{eqnarray}
\centering
\mathbf{F}_t=\frac{1}{H \times W} \sum_{i=1}^{H} \sum_{j=1}^{W}\mathrm{W}(\mathrm{Cat}({\mathbf{M}_1,\cdots,\mathbf{M}_4}) )(i,j),
\label{eq:feature}
\end{eqnarray}
where $\mathbf{F}_t \in \mathbb{R}^{{V} \times {C} \times {1} \times {1}}$ denotes the fused features; $\mathrm{Cat}(\cdot)$ denotes the concatenation operation; $\mathrm{W}(\cdot)$ denotes the $1\times1$ convolutional layer. To achieve precise and adaptive selection guidance, we create a weight vector $ \mathbf{V}_w \in \mathbb{R}^{{4} \times 1}$ and a selection vector $ \mathbf{V}_s \in \mathbb{R}^{{15} \times 1}$ using $\{\mathbf{M}_m\}_{m=1}^4$. Specifically, we flatten $\mathbf{F}_t$ in $(V, C)$ dimension, and apply two fully connected (FC) layers and softmax functions to obtain $\mathbf{V}_w$ and $\mathbf{V}_s$, which are responsible for weighting $\{\mathbf{M}_m\}_{m=1}^4$ and selecting their different combinations for following varied tasks.

For $\mathbf{V}_w$, to make $\{\mathbf{M}_m\}_{m=1}^4$ receive different attentions, it is calculated by multiplying $\mathbf{V}_w$ with corresponding position of $\{\mathbf{M}_m\}_{m=1}^4$. Since the combinations of $\{\mathbf{M}_m\}_{m=1}^4$ have 15 possibilities ($C_{4}^{1}$ + $C_{4}^{2}$ + $C_{4}^{3}$ + $C_{4}^{4}$ = 15), we adopt slightly different settings in last FC layer for getting $\mathbf{V}_s$, \ie, set the output channel number of last FC layer to 15. Finally, the proposed feature selection module is capable of dynamically selecting most appropriate features $\mathbf{S} \in \mathbb{R}^{{V} \times C \times \frac{H}{8} \times \frac{W}{8}}$ for different tasks. The whole process can be described as
\begin{eqnarray}
\centering
\mathbf{S}=\mathrm{Argmax}(\mathbf{V}_s\mid (\mathrm{W}(\mathrm{Cat}(\mathbf{C}_{\{\mathbf{M}_m\}_{m=1}^4\odot \mathbf{V}_w}^i)_{i=1}^4 )) ),
\label{eq:feature}
\end{eqnarray}
where $\odot$ denotes the element-wise multiplication; $\mathbf{C}_n^i$ denotes the permutation and combination operation. Finally, we can obtain $\mathbf{S}^r$, $\mathbf{S}^t$, $\mathbf{S}^d$, and $\mathbf{S}^q$ as inputs for different tasks.

\subsection{Optional Auxiliary Networks}

Here, we design three optional auxiliary networks to assist the main task. These optional networks are used to predict distortion range, distortion type, and distortion degree, respectively. The details are as follows.

\subsubsection{Distortion Range Prediction}

\begin{figure}[t]
\centering
\subfigure[The viewport distortion perception module.]{
    \includegraphics[width=1\linewidth]{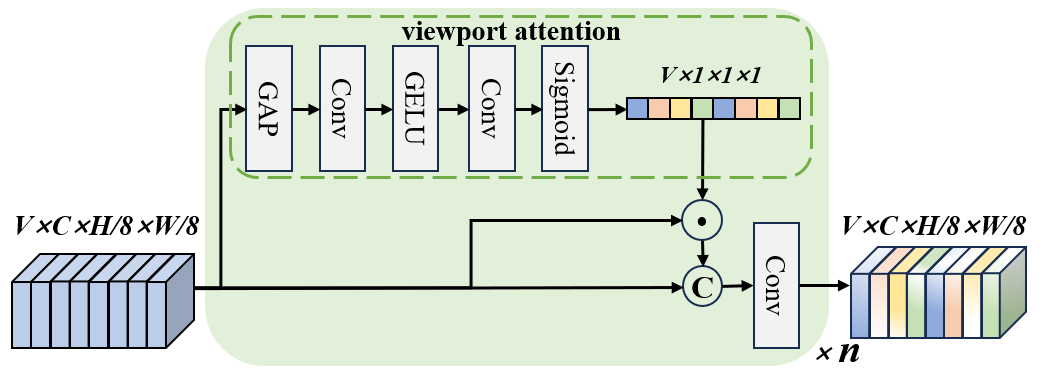}
}
\subfigure[The spatial distortion measurement module.]{
    \includegraphics[width=1\linewidth]{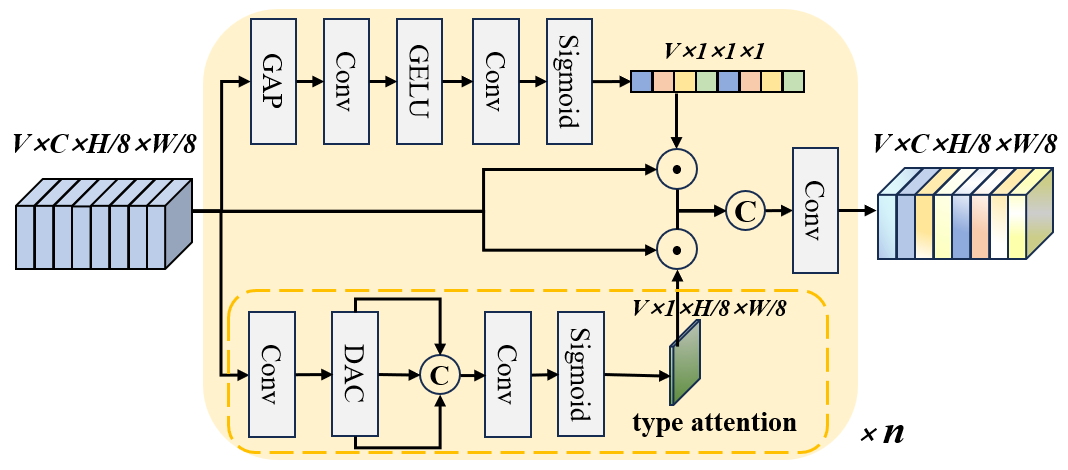}
}
\subfigure[The channel distortion computation module.]{
    \includegraphics[width=1\linewidth]{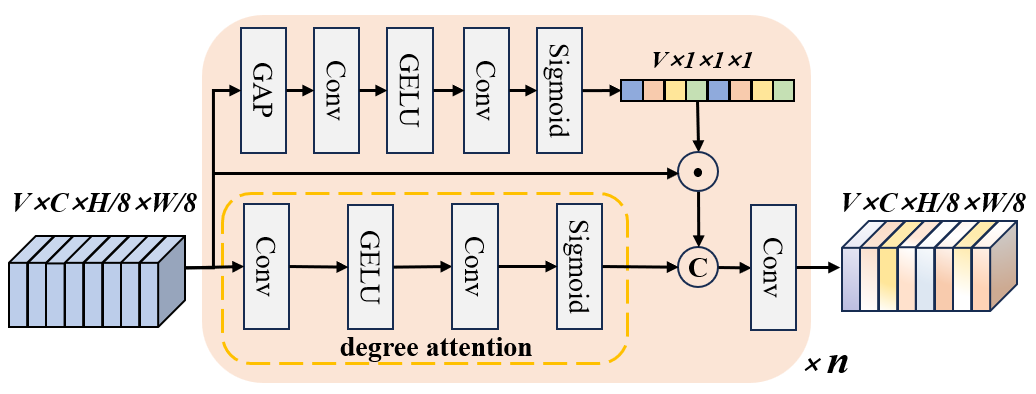}
}
\caption{The architecture of the core modules in three optional auxiliary networks.}
\label{fig:Auxiliary network module} 
\end{figure}

Compared to uniformly distorted OIs, non-uniformly distorted OIs show varied quality in single or multiple areas, making it necessary to identify the overall degree of distortion. Inspired by the effectiveness of attention mechanism~\cite{hu2018squeeze},~\cite{woo2018cbam},~\cite{shen2023blind}, we design a viewport distortion perception (VDP) module to capture the quality coherence among the viewport sequence, whose structure is illustrated in Fig.~\ref{fig:Auxiliary network module} (a). Specifically, its input $\mathbf{S}^r$ is processed by a viewport attention (VA) block, in which the global information is firstly embedded by a GAP operation, then two $1\times1$ convolutional layers are used to capture the viewport dependency vector $\mathbf{X}_v \in \mathbb{R}^{{V} \times {1} \times {1} \times {1}}$:
\begin{eqnarray}
\begin{aligned}
\mathbf{X}_v &= \mathrm{VA}(\mathbf{S}^r) \\
&=\sigma (\mathrm{W}^2(\delta(\mathrm{W}^1(\mathcal{F}_{gap}(\mathbf{S}^r)))),
\label{eq:feature}
\end{aligned}
\end{eqnarray}
where $\mathcal{F}_{gap}$ denotes the GAP operation; $\delta$ denotes the gaussian error linear unit (GELU) function; $\sigma$ refers to the sigmoid function; the kernel sizes of $\mathrm{W}^1$ and $\mathrm{W}^2$ are ${128} \times {64} \times {1} \times {1}$ and ${64} \times {1} \times {1} \times {1}$, respectively. By interacting $\mathbf{S}^r$ and $\mathbf{X}_v$ in the $V$ dimension, the distortion range information can be retained and highlighted. The whole process of the VDP module can be described as
\begin{eqnarray}
\begin{aligned}
\mathbf{VP} &= \mathrm{VDP}_n(\mathbf{S}^r) \\
&= \mathrm{W}(\mathrm{Cat}(\mathbf{X}_v \odot \mathbf{S}^r,\mathbf{S}^r)),
\end{aligned}
\label{eq:feature}
\end{eqnarray}
where $\mathbf{VP} \in \mathbb{R}^{{V} \times {C} \times \frac{H}{8} \times \frac{W}{8}}$ represents the resultant features; $\mathrm{VDP}_n(\cdot)$ denotes the VDP module and can be applied $n$ times. Finally, a non-linear mapping, including a GAP operation, a transposition operation, and an FC layer, is used to predict the probability vector of the distortion range $\mathbf{M}_{rcm} \in \mathbb{R}^{{1} \times {r}}$, where $r$ denotes the number of distortion range. The process can be described as
\begin{eqnarray}
\centering
\mathbf{M}_{rcm} = \mathrm{Softmax}(\mathrm{MLP}(\mathrm{T}(\mathcal{F}_{gap}(\mathbf{V})))).
\label{eq:6}
\end{eqnarray}

\subsubsection{Distortion Type Prediction}

Here, we design a spatial distortion measurement (SDM) module, whose architecture is shown in Fig.~\ref{fig:Auxiliary network module} (b). Specifically, the SDM module has two pathways named the VA block and the type attention (TA) block. In the bottom pathway, \ie, the TA block, a 1$\times$1 convolution is used to compress the input $\mathbf{S}^t$ in channel dimension to 1, and then three dilated convolutions $\mathrm{DAC_{3\times 3}^3}$, $\mathrm{DAC_{5\times 5}^5}$, $\mathrm{DAC_{3\times 3}^1}$ with kernels $3\times3$, $5\times5$, $3\times3$ and dilation rates 3, 5, 1 are used to capture multiscale spatial distortion information without reducing spatial resolution. By doing this, we can obtain three outputs $\mathbf{S}^{t_1}$, $\mathbf{S}^{t_2}$, and $\mathbf{S}^{t_3}$ with different scales. Subsequently, $\mathbf{S}^{t_1 }$, $\mathbf{S}^{t_2}$, and $\mathbf{S}^{t_3}$ are fused by a $1\times1$ convolution. We then apply the sigmoid function to get the spatial weight map $\mathbf{X}_s \in \mathbb{R}^{{V} \times {1} \times \frac{H}{8} \times \frac{W}{8}}$:
\begin{eqnarray}
\begin{aligned}
\mathbf{X}_s &= \mathrm{TA}(\mathbf{S}^t) \\
&= \sigma(\mathrm{W}(\mathrm{Cat}(\mathbf{S}^{t_1 },\mathbf{S}^{t_2},\mathbf{S}^{t_3 }))),
\end{aligned}
\label{eq:feature}
\end{eqnarray}

Subsequently, $\mathbf{X}_s$ and $\mathbf{S}^t$ interact and fuse in $(V, H, W)$ dimensions, obtaining the spatial perception feature $\mathbf{T} \in \mathbb{R}^{{V} \times {C} \times \frac{H}{8} \times \frac{W}{8}}$. The process of SDM can be described as
\begin{eqnarray}
\begin{aligned}
\mathbf{T} &= \mathrm{SDM}_n(\mathbf{S}^t) \\
&= \mathrm{W}(\mathrm{Cat}(\mathrm{TA}(\mathbf{S}^t) \odot \mathbf{S}^t, \mathrm{VA}(\mathbf{S}^t) \odot \mathbf{S}^t)),
\end{aligned}
\label{eq:feature}
\end{eqnarray}
where $\mathrm{SDM}_n(\cdot)$ refers to the SDM module that can be executed $n$ times. Finally, the same as $\mathbf{M}_{rcm}$, the type classification map $\mathbf{M}_{tcm} \in \mathbb{R}^{{1} \times {t}}$ is obtained by equation~\ref{eq:6}, where $t$ denotes the number of distortion type.

\subsubsection{Distortion Degree Prediction}
Compared to distortion range prediction and distortion type prediction tasks, the distortion degree of an OI is directly related to its quality, and therefore it is relatively easy to recognize distortion degree. To this end, we design a channel distortion computation (CDC) module to help the model understand the impact of distortion degree, whose architecture is shown in Fig.~\ref{fig:Auxiliary network module} (c). Specifically, we use $\mathbf{S}^d$ as input and apply a $1\times1$ convolutional layer to make its number of channels half the original. After introducing the nonlinear relationship through the GELU function, the number of channels of $\mathbf{S}^d$ is transformed to the original number of channels through the $1\times1$ convolutional layer and the channel attention feature $\mathbf{X}_d \in \mathbb{R}^{{V} \times {C} \times \frac{H}{8} \times \frac{W}{8}} $ is obtained after the sigmoid function. This process can be formulated as 
\begin{eqnarray}
\begin{aligned}
\mathbf{X}_d &= \mathrm{DA}(\mathbf{S}^d) \\
&=\sigma (\mathrm{W}^3(\delta(\mathrm{W}^1(\mathbf{S}^d))),
\end{aligned}
\label{eq:feature}
\end{eqnarray}
where the kernel sizes of $\mathrm{W}^3$ is ${64} \times {128} \times {1} \times {1}$. Due to the characteristics of non-uniform distortion, we parallel the VA block and DA block to capture the local distortion severity. Then, $\mathbf{X}_d$ and $\mathbf{S}^d$ are interacted in $(V, C)$ dimension to obtain the distortion degree feature $\mathbf{D} \in \mathbb{R}^{{V} \times {C} \times\frac{H}{8} \times \frac{W}{8}}$. The computational process of the CDC module can be formulated as
\begin{eqnarray}
\begin{aligned}
\mathbf{D} &= \mathrm{CDC}_n(\mathbf{S}^d) \\
&= \mathrm{W}(\mathrm{Cat}(\mathrm{DA}(\mathbf{S}^d), \mathrm{VA}(\mathbf{S}^d) \odot \mathbf{S}^d),
\end{aligned}
\label{eq:feature}
\end{eqnarray}
where $\mathrm{SDM}_n(\cdot)$ means the SDM module and can be repeated $n$ times. The same as $\mathbf{M}_{rcm}$, the degree classification map $\mathbf{M}_{dcm} \in \mathbb{R}^{{1} \times {d}}$ is obtained through equation 8, where $d$ denotes the number of distortion degree.

\subsection{Main Task Sub-Network}

\begin{figure}[t]
\centering
\includegraphics[width=1\linewidth]{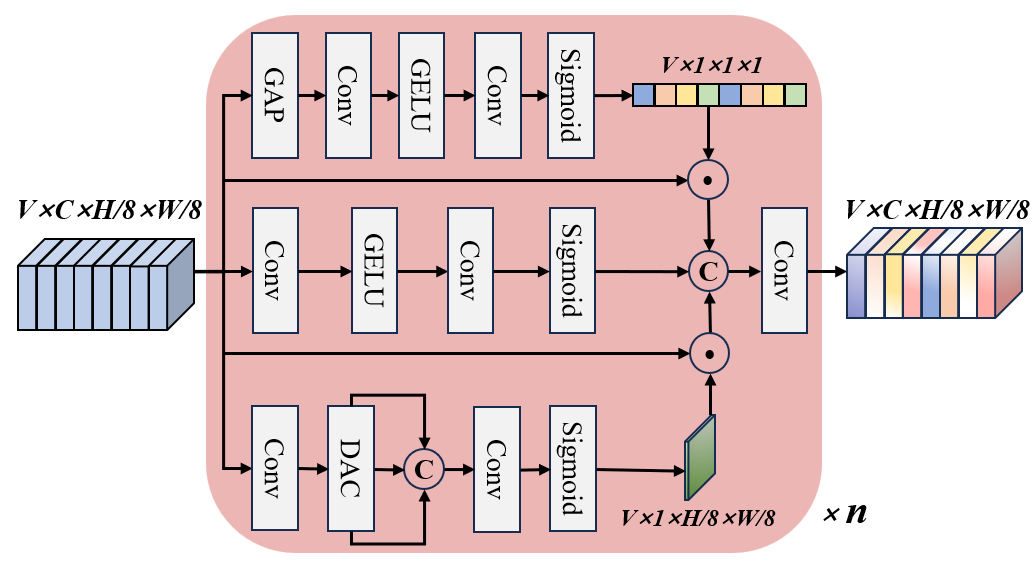}
\caption{The multidimension distortion integration module.}
\label{fig:MDI}
\end{figure}

\begin{figure}[t]
\centering
\includegraphics[width=1\linewidth]{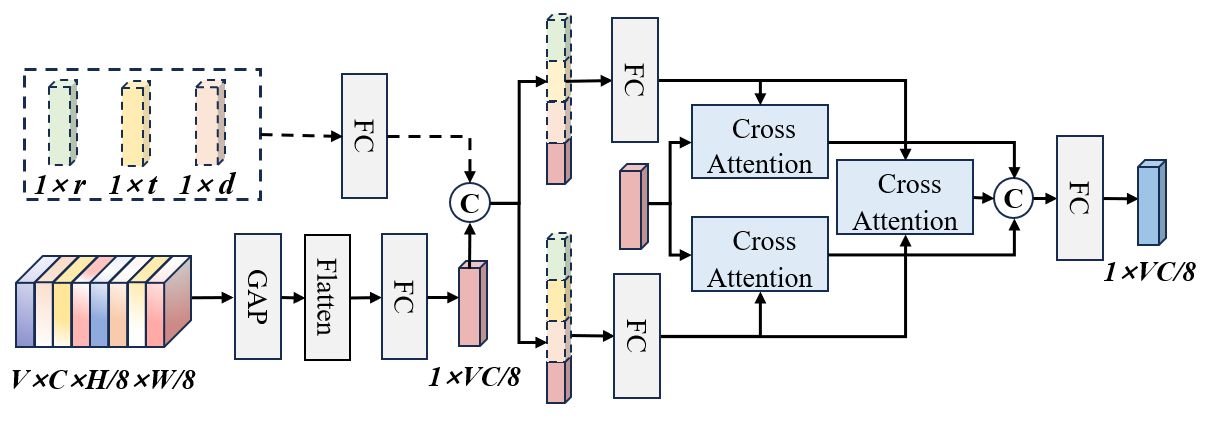}
\caption{The multitask auxiliary fusion module. Note that the dashed line indicates that this part becomes unactivated when the auxiliary task is not applicable.}
\label{fig:maf}
\end{figure}

This sub-network consists of a multidimension distortion integration (MDI) module and a multitask auxiliary fusion (MAF) module, as shown in Figs.~\ref{fig:MDI} and \ref{fig:maf}. For the MDI module, the input $\mathbf{S}^q$ is processed by three parallel modules, \ie, a VA block, a TA block, and a DA block, and then their outputs are concatenated and the multidimensional feature representation $\mathbf{Q} \in \mathbb{R}^{{V} \times {C} \times\frac{H}{8} \times \frac{W}{8}}$ is obtained. This process can be described as 
\begin{eqnarray}
\begin{aligned}
\mathbf{Q} &= \mathrm{MDI}_n(\mathbf{S}^q) \\
&= \mathrm{W}(\mathrm{Cat}(\mathrm{VA}(\mathbf{S}^q) \odot \mathbf{S}^q, \mathrm{TA}(\mathbf{S}^q) \odot \mathbf{S}^q, \mathrm{DA}(\mathbf{S}^q)),
\end{aligned}
\label{eq:feature}
\end{eqnarray}
where $\mathrm{MDI}_n(\cdot)$ means that the MDI module can be repeated $n$ times.

For the MAF module, the representation $\mathbf{Q}$ is first transformed by serialized GAP, flatten, and FC operations, the output of which is denoted by $\mathbf{F} \in \mathbb{R}^{{1} \times \frac{VC}{8}}$. Then, for the fusion of different semantic information and computational convenience, we map the dimensions $r$, $t$, and $d$ in $\mathbf{M}_{rcm}$, $\mathbf{M}_{tcm}$, and $\mathbf{M}_{dcm}$ to $C$, and concatenate them through a linear layer to obtain cross-semantic feature $\mathbf{C} \in \mathbb{R}^{{1} \times C}$. Subsequently, to allow semantic information from different auxiliary tasks to fully interact with the main task, we use cross-attention~\cite{gheini2021cross} to capture interactive information $\Bar{\mathbf{I}} \in \mathbb{R}^{{1} \times {3C}}$:
\begin{eqnarray}
\centering
\left\{\begin{matrix}
\Bar{\mathbf{I}} = \mathrm{Cat}(\mathrm{CA}(\mathbf{C},\mathbf{F}),\mathrm{CA}(\mathbf{F}, \mathbf{C}),\mathrm{CA}(\mathbf{C},\mathbf{C})),
   
 \\\mathrm{CA}(\mathbf{X}_1, \mathbf{X}_2) = \mathrm{Softmax}\left(\frac{\mathbf{Q} \cdot  \mathbf{K}^\mathrm{T}}{\sqrt{d_2}}\right) \cdot \mathbf{V},
 \\ \mathbf{Q} = \mathbf{X}_1\cdot \mathbf{W}_Q; \mathbf{K} = \mathbf{V} =\mathbf{X}_2\cdot \mathbf{W}_K,

\end{matrix}\right.
\label{eq:feature}
\end{eqnarray}
where $\mathrm{CA}(\cdot)$ denotes the cross-attention module; $\mathbf{X}_1$ and $\mathbf{X}_2$ are two inputs of the CA module; $d_2$ is the dimension of $\mathbf{X}_2$; $\mathbf{W}_Q$ and $\mathbf{W}_K$ are the weight matrices; To avoid too many parameters, $\Bar{\mathbf{I}}$ becomes the interaction feature $\mathbf{I} \in \mathbb{R}^{{1} \times {C}}$ through FC layer. Finally, we can obtain the predicted perceptual quality score $\hat{s}$ for an OI by a FC layer.

\subsection{Loss Function}
To train the main task and optional auxiliary tasks jointly, each task needs to choose an appropriate loss function to update the network parameters. We use the mean squared error (MSE) loss function to optimize the main network:
\begin{eqnarray}
\centering 
\mathrm{L_1}=\frac{1}{N} \sum_{i=1}^{N} (\hat{s}_i - s_i)^2,
\label{eq:feature}
\end{eqnarray}
where $\hat{s}_i$ and $s_i$ are the predicted quality score and the subjective score of the $i$-th OI, respectively.
We adopt the cross entropy loss function to optimize auxiliary networks:
\begin{eqnarray}
\centering 
\mathrm{L_2}=-\sum_{i=1}^{N} y_i \cdot log{\hat{y_i}},
\label{eq:feature}
\end{eqnarray}
where $\hat{y}_i$ and $y_i$ are the predicted distortion label and the true distortion label of the $i$-th OI, respectively.

Considering that the weights of different task loss functions will affect the performance of the multitask learning model, we introduce uncertainty weighting~\cite{kendall2018multi} to automatically determine the weight size of different loss functions:
\begin{eqnarray}
\centering 
L(\sigma_1,\cdots,\sigma_k ) = \frac{1 }{2\sigma_1^2}L_1 + \ln{\sigma_1}+\cdots +\frac{1 }{2\sigma_k^2}L_2 + \ln{\sigma_k},
\label{eq:feature}
\end{eqnarray}
where $\sigma$ is the training weight for different loss functions; $k$ is the number of auxiliary tasks and ranges from 1 to 3.

\section{Experiments and results}
\label{sec:qu_ex}

In this section, we first introduce the implementation details of our experiments, test databases, and evaluation metrics, then compare the proposed model with other state-of-the-art methods on the test databases. Finally, we verify the effectiveness of the proposed modules.

\subsection{Implementation Details}

\paragraph{Experimental Settings}
The input viewport size is fixed to 3$\times$224$\times$224. The numbers of $V$ and $C$ are 8 and 128, respectively. The number of channels $C1$, $C2$, $C3$, and $C4$ are 256, 512, 1024, and 1024. The number of $n$ for VDP, SDM, CDC, and MDI modules is 8, 4, 4, and 4. The model is optimized using the adaptive moment estimation (Adam) optimizer with an initial learning rate of $10^{-5}$. We randomly split $80\%$ OIs of each database for training and the remaining $20\%$ for testing. The batch size is set to 8 and the training process ends after 50 epochs. Our experiments are implemented on a high-performance server with Intel(R) Xeon(R) Gold 6326 CPU@2.90GHz, 24G NVIDIA GeForce RTX A5000 GPU, and 260GB RAM. 

\paragraph{Evaluation Criteria}
We employ three common evaluation metrics to quantify the quality prediction performance, including Pearson's linear correlation coefficient (PLCC), Spearman's rank order correlation coefficient (SRCC), and root mean square error (RMSE). The values of PLCC and SRCC range from 0 to 1, a higher value indicates better performance and a lower RMSE indicates a more accurate prediction. Following the suggestion in~\cite{vqeg}, a four-parameter logistic function is applied before calculating PLCC and RMSE as follows:
\begin{eqnarray}
\centering 
f(\hat{s})=\frac{\eta_1-\eta_2}{1+e^{(-\frac{\hat{s}-\eta_3}{\eta_4})}}+\eta_2,
\label{eq:nonlinear_mapping}
\end{eqnarray}
where $\left\{\eta_i|i=1,\cdots,4\right\}$ are parameters to be fitted.

\begin{figure}[t]
\centering
\includegraphics[width=0.9\linewidth]{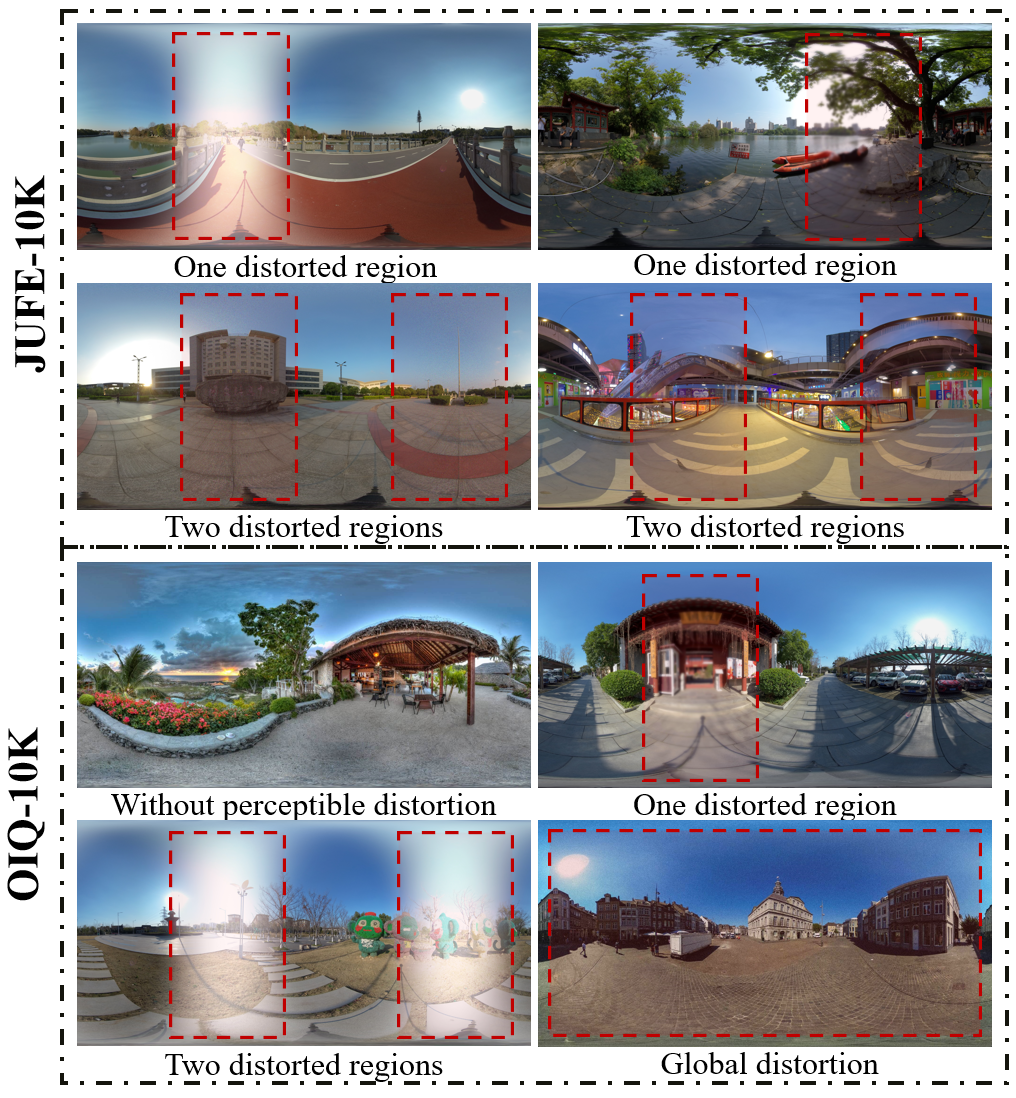}
\caption{Visual examples of OIs from the JUFE-10K and OIQ-10K databases. The distorted region is highlighted by a red box.}
\label{fig:database}
\end{figure}

\paragraph{Test Databases}
We utilize two benchmark OIQA databases in the experiments, \ie, JUFE-10K~\cite{JUFE-10K} and OIQ-10K~\cite{OIQ-10K}. The \textbf{JUFE-10K}\footnote{https://github.com/RJL2000/OIQAND} database is constructed to simulate those in-wild OIs that are easily disturbed by a variety of non-uniform distortions. It includes 10,320 non-uniformly distorted OIs generated from 430 reference OIs with 1 or 2 distorted region(s), 4 distortion types, and 3 distortion levels. Distortion types include Gaussian noise (GN), Gaussian blur (GB), brightness discontinuities (BD), and stitching distortion (ST). Distortion ranges include one distorted region and two distorted regions. Subjective quality scores range from 1 to 5. Note that a higher score means better visual quality. The \textbf{OIQ-10K}\footnote{https://github.com/WenJuing/IQCaption360} database considers both uniform and non-uniform distortion scenarios. It includes 10,000 OIs with different distortion ranges, which include `without perceptibly distorted region', `one distorted region', `two distorted regions', and `global distortion'. And it also provides the head and eye movement data. The subjective quality scores are in the range of 1 to 3. Some visual examples from these two databases are shown in Fig.~\ref{fig:database}.

\begin{table*}
\caption{Performance comparison of state-of-the-art BIQA and OIQA methods on the JUFE-10K database. * represents that MTAOIQA is not equipped with any auxiliary tasks. The best results are highlighted in bold.}
\begin{tabular}{m{0.8cm}<{\centering} m{2.5cm}<{\centering}
 m{0.45cm}<{\centering} m{0.45cm}<{\centering} m{0.7cm}<{\centering} m{0.45cm}<{\centering} m{0.45cm}<{\centering} m{0.7cm}<{\centering} m{0.45cm}<{\centering} m{0.45cm}<{\centering} m{0.7cm}<{\centering} m{0.45cm}<{\centering} m{0.45cm}<{\centering} m{0.7cm}<{\centering} m{0.45cm}<{\centering} m{0.45cm}<{\centering} m{0.7cm}<{\centering} }
\toprule
\multirow{4}{*}{\centering Type} & \multirow{4}{*}{Models} & \multicolumn{3}{c}{BD}  & \multicolumn{3}{c}{GB}   & \multicolumn{3}{c}{GN}   & \multicolumn{3}{c}{ST}   & \multicolumn{3}{c}{Overall} \\ \cmidrule(lr){3-5}\cmidrule(lr){6-8}\cmidrule(lr){9-11}\cmidrule(lr){12-14}\cmidrule(lr){15-17}
                      &   & PLCC & SRCC  & RMSE & PLCC & SRCC & RMSE & PLCC & SRCC & RMSE & PLCC & SRCC & RMSE & PLCC & SRCC & RMSE \\
\midrule
\multirow{8}{*}{\makecell[c]{BIQA}} & NIQE~\cite{mittal2012making} & 0.226 & 0.199 & 0.602 & 0.147 & 0.147 & 0.596 & 0.113 & 0.047 & 0.517 & 0.139 & 0.090 & 0.554 & 0.094 & 0.047 & 0.605\\[2pt]
                      & HyperIQA~\cite{su2020blindly} & 0.267 & 0.265 & 0.595 & 0.361 & 0.353 & 0.561 & 0.256 & 0.252 & 0.502 & 0.122 & 0.102 & 0.555 & 0.201 & 0.198 & 0.595\\[2pt]
                      & UNIQUE~\cite{zhang2021uncertainty} & 0.209 & 0.204 & 0.604 & 0.285 & 0.277 & 0.577 & 0.380 & 0.374 & 0.481 & 0.123 & 0.109 & 0.555 & 0.174 & 0.169 & 0.599 \\[2pt]
                      & CONTRIQUE~\cite{madhusudana2022image} & 0.168 & 0.157 & 0.609 & 0.036 & 0.029 & 0.602 & 0.050 & 0.051 & 0.519 & 0.052 & 0.016 & 0.559 & 0.089 & 0.076 & 0.606\\[2pt]
                      & MANIQA~\cite{yang2022maniqa} & 0.009 & 0.008 & 0.618 & 0.126 & 0.109 & 0.598 & 0.287  &  0.279  & 0.48 & 0.124 & 0.073 & 0.555 & 0.101 & 0.090 &  0.605\\[2pt]
                      & VCRNet~\cite{pan2022vcrnet} & 0.222 & 0.214 & 0.602 & 0.339 & 0.340 & 0.566 & 0.092  &  0.094  & 0.517 &
                      0.052 & 0.042 & 0.558 & 0.162  &  0.150 &  0.599\\[2pt]
                      & LIQE~\cite{zhang2023blind} & 0.031 & 0.028 &  0.618 & 0.036 & 0.039 & 0.601 & 0.494  &  0.481  & 0.452 & 0.105 & 0.005 & 0.556 & 0.029  &  0.026 &  0.607\\[2pt]
                      & SAAN~\cite{yi2023towards} & 0.184 & 0.170 & 0.607 & 0.095 & 0.088 & 0.599 & 0.001 &  0.013 & 0.519 & 0.087 & 0.087 & 0.557 & 0.056 & 0.040 & 0.607\\[2pt]
\midrule
\multirow{4}{*}{\makecell[c]{FR-\\OIQA}}   & S-PSNR~\cite{s-psnr}    & 0.692 & 0.694 & 0.446  & 0.413 & 0.276 & 0.549  & 0.448 & 0.275 & 0.465  & 0.152 & 0.130 & 0.553 & 0.355 &0.285 & 0.568\\[2pt]
                      & WS-PSNR~\cite{ws-psnr}   & 0.686 & 0.688 & 0.450 & 0.412 & 0.275 & 0.549  & 0.449 & 0.279 & 0.465 & 0.141 & 0.114 &0.554 & 0.353 & 0.284 &0.569\\[2pt]
                      & CPP-PSNR~\cite{cpp-psnr}  & 0.686 & 0.688 & 0.450  & 0.412 & 0.275 & 0.549 & 0.448 & 0.274 & 0.465 & 0.140 & 0.114 & 0.554 & 0.355 & 0.285 & 0.568  \\[2pt]
                      & WS-SSIM~\cite{ws-ssim} & 0.415 &  0.310 & 0.562 & 0.462  &  0.312  &  0.534 & 0.448 & 0.289 &   0.465 & 0.110 &0.055 & 0.556 & 0.388 & 0.249  & 0.560\\[2pt]
\midrule
\multirow{6}{*}{\makecell[c]{BOIQA}}   & MC360IQA~\cite{MC360IQA}   & 0.735 & 0.735 & 0.430  & 0.721 & 0.725 & 0.412  & 0.539 & 0.528 & 0.450  & 0.440 & 0.432 & 0.481  & 0.620 & 0.611 & 0.474\\[2pt]
  & VGCN~\cite{VGCN}    & 0.782 & 0.770 & 0.396  & 0.443 & 0.409 & 0.533  & 0.102 & 0.087 & 0.531  & 0.207 & 0.190 & 0.524 & 0.471  & 0.377 & 0.533 \\[2pt]
  & Fang22~\cite{fang2022perceptual}    & 0.759 & 0.753 & 0.413 & 0.725 & 0.729 & 0.409 & 0.535 & 0.539 & 0.451 & 0.390 & 0.386 & 0.494 & 0.633  & 0.616 & 0.468 \\[2pt]
  & Assessor360~\cite{wu2023assessor360}   & 0.727 & 0.726 & 0.435 & 0.688 & 0.696 & 0.431 & 0.733 & 0.747 & 0.363 & 0.518 & 0.516 & 0.458 & 0.694 & 0.690 & 0.435 \\[2pt]
    & MTAOIQA*   & \textbf{0.860} & \textbf{0.859} & \textbf{0.323}  & \textbf{0.813} &\textbf{0.815} & \textbf{0.346}  & \textbf{0.775} & \textbf{0.785} & \textbf{0.337}  & \textbf{0.651} & \textbf{0.650} & \textbf{0.407}  & \textbf{0.798}  & \textbf{0.798} & \textbf{0.364} \\[2pt]
     & MTAOIQA    & \textbf{0.875} & \textbf{0.875} & \textbf{0.307}  & \textbf{0.829} &\textbf{0.831} & \textbf{0.333}  & \textbf{0.786} & \textbf{0.794} & \textbf{0.330}  & \textbf{0.689} & \textbf{0.684} & \textbf{0.388}  & \textbf{0.822}  & \textbf{0.821} & \textbf{0.344} \\[2pt]
\bottomrule
\end{tabular}

\label{tab-JUFE-10K-DB}
\end{table*}

\begin{table*}
\caption{Performance comparison of state-of-the-art BIQA and OIQA metrics on the OIQ-10K database in detail. The abbreviations `R1', `R2', `R3', and `R4' denote `without perceptibly distorted region', `one distorted region', `two distorted regions', and `global distortion', respectively. * represents that MTAOIQA is not equipped with any auxiliary tasks. The best results are marked in bold.}
\begin{tabular}{m{0.8cm}<{\centering} m{2.5cm}<{\centering}
 m{0.45cm}<{\centering} m{0.45cm}<{\centering} m{0.7cm}<{\centering} m{0.45cm}<{\centering} m{0.45cm}<{\centering} m{0.7cm}<{\centering} m{0.45cm}<{\centering} m{0.45cm}<{\centering} m{0.7cm}<{\centering} m{0.45cm}<{\centering} m{0.45cm}<{\centering} m{0.7cm}<{\centering} m{0.45cm}<{\centering} m{0.45cm}<{\centering} m{0.7cm}<{\centering} }
\toprule
\multirow{4}{*}{\centering Type} & \multirow{4}{*}{Models} & \multicolumn{3}{c}{R1}  & \multicolumn{3}{c}{R2}   & \multicolumn{3}{c}{R3}   & \multicolumn{3}{c}{R4}   & \multicolumn{3}{c}{Overall} \\ \cmidrule(lr){3-5}\cmidrule(lr){6-8}\cmidrule(lr){9-11}\cmidrule(lr){12-14}\cmidrule(lr){15-17}
                      &   & PLCC & SRCC  & RMSE & PLCC & SRCC & RMSE & PLCC & SRCC & RMSE & PLCC & SRCC & RMSE & PLCC & SRCC & RMSE \\
\midrule
\multirow{8}{*}{\makecell[c]{BIQA}} & NIQE~\cite{mittal2012making} & 0.178 & 0.123 & 0.430 & 0.075 & 0.080 & 0.324 & 0.098 & 0.067 & 0.363 & 0.400 & 0.417 & 0.421 & 0.368 & 0.271 & 0.436\\[2pt]
                      & HyperIQA~\cite{su2020blindly} & 0.265 & 0.251 & 0.421 & 0.164 & 0.160 & 0.321 & 0.242 & 0.239 & 0.354 & 0.581 & 0.577 & 0.374 & 0.475 & 0.444 & 0.412\\[2pt]
                      & UNIQUE~\cite{zhang2021uncertainty} & 0.408 & 0.399 & 0.399 & 0.187 & 0.183 & 0.319 & 0.242 & 0.238 & 0.354 & 0.663 & 0.667 & 0.344 & 0.573 & 0.538 & 0.384 \\[2pt]
                      & CONTRIQUE~\cite{madhusudana2022image} & 0.210 & 0.122 & 0.427 & 0.149 & 0.142 & 0.322 & 0.219 & 0.207 & 0.356 & 0.417 & 0.422 & 0.420 & 0.309 & 0.263 & 0.446\\[2pt]
                      & MANIQA~\cite{yang2022maniqa} & 0.028 & 0.034 & 0.357 & 0.004 & 0.018 & 0.363 & 0.164 & 0.155 & 0.458 & 0.296 & 0.296 & 0.416 & 0.113 & 0.098 & 0.466\\[2pt]
                      & VCRNet~\cite{pan2022vcrnet} & 0.187 & 0.182 & 0.429 & 0.082 & 0.070 & 0.324 & 0.242 & 0.230 & 0.354 & 0.341 & 0.340 & 0.432 & 0.289 & 0.277 & 0.449\\[2pt]
                      & LIQE~\cite{zhang2023blind} & 0.637 & 0.651 & 0.336 & 0.090 & 0.081 & 0.324 & 0.075 & 0.065 & 0.364 & 0.683 & 0.623 & 0.336 & 0.551 & 0.522 & 0.391\\[2pt]
                      & SAAN~\cite{yi2023towards} & 0.062 & 0.056 & 0.436 & 0.056 & 0.058 & 0.325 & 0.133 & 0.109 & 0.362 & 0.130 & 0.035 & 0.456 & 0.148 & 0.036 & 0.464\\[2pt]
\midrule
\multirow{4}{*}{\makecell[c]{FR-\\OIQA}}   & S-PSNR~\cite{s-psnr}    & - & - & -  & 0.237 & 0.216 &                         0.316  & 0.359 & 0.275 & 0.341  & - & - & - & 0.302 & 0.252 & 0.343\\[2pt]
                      & WS-PSNR~\cite{ws-psnr}   & - & - & - & 0.220 & 0.188 & 0.317 & 0.355 & 0.271 & 0.341 & - & - & - & 0.295 & 0.248 & 0.344\\[2pt]
                      & CPP-PSNR~\cite{cpp-psnr}  & - & - & - & 0.220 & 0.188 & 0.317 & 0.355 & 0.271 & 0.341 & - & - & - & 0.295 & 0.248 & 0.344  \\[2pt]
                      & WS-SSIM~\cite{ws-ssim} & - &  - & - & 0.076 & 0.069 & 0.324 & 0.259 & 0.094 &  0.353 & - & - & - & 0.223 & 0.062  & 0.351\\[2pt]
\midrule
\multirow{6}{*}{\makecell[c]{BOIQA}}   & MC360IQA~\cite{MC360IQA}   & 0.502 & 0.462 & 0.386  & 0.446 & 0.424 & 0.273  & 0.626 & 0.625 & 0.271  & 0.782 & 0.767 & 0.281  & 0.721 & 0.710 & 0.319\\[2pt]
  & VGCN~\cite{VGCN}    & 0.411 & 0.383 & 0.407  & 0.498 & 0.479 & 0.265  & 0.654 & 0.649 & 0.263  & 0.763 & 0.728 & 0.292  & 0.706  & 0.699 & 0.325 \\[2pt]
  & Fang22~\cite{fang2022perceptual}    & 0.553 & 0.531 & 0.372  & 0.542 & 0.522 & 0.257  & 0.673 & 0.670 & 0.257  & 0.832 & 0.798 & 0.251  & 0.769  & 0.758 & 0.293 \\[2pt]
  & Assessor360~\cite{wu2023assessor360}   & 0.678 & 0.679 & 0.328 & 0.508 & 0.479 & 0.263 & 0.663 & 0.652 & 0.260 & 0.838 & 0.811 & 0.247 & 0.790 & 0.773 & 0.281 \\[2pt]
    & MTAOIQA*    & \textbf{0.717} & \textbf{0.715} & \textbf{0.311}  & \textbf{0.598} &\textbf{0.595} & \textbf{0.245}  & \textbf{0.716} & \textbf{0.723} & \textbf{0.243}  & \textbf{0.847} & \textbf{0.817} & \textbf{0.240}  & \textbf{0.819}  & \textbf{0.813} & \textbf{0.263} \\[2pt]
     & MTAOIQA    & \textbf{0.717} & \textbf{0.719} & \textbf{0.311}  & \textbf{0.660} &\textbf{0.653} & \textbf{0.229}  & \textbf{0.731} & \textbf{0.734} & \textbf{0.237}  & \textbf{0.861} & \textbf{0.831} & \textbf{0.230}  & \textbf{0.829}  & \textbf{0.824} & \textbf{0.256} \\[2pt]
\bottomrule
\end{tabular}

\label{tab-OIQ-10K-DB}
\end{table*}

\subsection{Performance Comparison}
To prove the effectiveness of the proposed model, we compare the performance of MTAOIQA with eight BIQA methods: NIQE~\cite{mittal2012making}, HyperIQA~\cite{su2020blindly}, UNIQUE~\cite{zhang2021uncertainty}, CONTRIQUE~\cite{madhusudana2022image}, MANIQA~\cite{yang2022maniqa}, VCRNet~\cite{pan2022vcrnet}, LIQE~\cite{zhang2023blind}, and SAAN~\cite{yi2023towards}; Four FR-OIQA methods: S-PSNR~\cite{s-psnr}, WS-PSNR~\cite{ws-psnr}, CPP-PSNR~\cite{cpp-psnr}, and WS-SSIM~\cite{ws-ssim}; and four BOIQA methods: MC360IQA~\cite{MC360IQA}, VGCN~\cite{VGCN}, Fang22~\cite{fang2022perceptual}, and Assessor360~\cite{wu2023assessor360}. For a fair comparison, we retrain BOIQA methods using the same configuration. The experimental results on JUFE-10K and OIQ-10K are listed in Table~\ref{tab-JUFE-10K-DB} and Table~\ref{tab-OIQ-10K-DB}, where we can clearly observe that the results of the BIQA and FR-OIQA models are significantly inferior to BOIQA models in terms of overall performance. This is not surprising since there is a {natural gap between 2D images and OIs, which makes the BIQA model not applicable to OIs. HyperIQA and UNIQUE perform well compared to other BIQA models on the JUFE-10K and OIQ-10K databases, which is attributed to the multi-scale feature extraction and fusion in HyperIQA facilitate the model's understanding of local distortion and global information, and the training strategy for multiple databases and an uncertainty regularizer contribute to UNIQUE's generalization ability. It is worth noting that LIQE performs quite differently on these two databases, the possible reason is that LIQE performs random cropping operations on test images, which may result in the cropped patches being with visible distortion or without distortion in the same OI. Therefore, it performs poorly on the JUFE-10K database with only non-uniform distortion and performs well on the OIQ-10K database with multiple distortion ranges. The unsatisfactory performance of SAAN is due to that the aesthetic features extracted by SAAN are not generic in the IQA task. These FR-OIQA methods consider the spherical representation of OIs, which alleviates part of the projection distortion. In contrast, the BOIQA methods, including MC360IQA, VGCN, Fang22, and Assessor360 are all viewport based, by simulating the characteristics of human real viewing OI content. This allows the model to correlate with subjective ratings and improve the accuracy of the predictions.

In terms of individual distortion, the performance on the BD, GB, and GN distortion types is higher than that on the ST distortion type, the reason of which is that the ST distortion exhibits structure inconsistencies and therefore makes the model undervalue the quality of OIs. Moreover, in Table~\ref{tab-OIQ-10K-DB}, we can observe that the results on the R2 case are worse than those of the other cases, which further indicates that the non-uniform distortion is challenging for the existing models. The performance of MC360IQA on the BD, GN, and GB distortion is significantly better than that on the ST distortion, which is attributed to the fact that MC360IQA over-samples viewports by adjusting the sampling starting point and thus results in that the ST distortion is easier being overlooked. VGCN performs well on the BD distortion, but the predictions for the other types are not satisfactory. The reason may be that the interaction information between the viewports in the local branch is fixed, which leads to greater dependence on the global feature of distorted OIs in ERP format. Fang22 shows better results than other BOIQA methods on the R2 and R3 cases, probably due to the fusing operation of viewing conditions. We can observe that Assessor360 performs suboptimal results in the R1 and R4 cases, which may benefit from the design of recursive probability viewport sampling and the temporal modeling processing of multi-viewport sequences. From Tables~\ref{tab-JUFE-10K-DB} and \ref{tab-OIQ-10K-DB}, we can observe that the proposed MTAOIQA and MTAOIQA* outperform the state-of-the-art models on these two databases. Compared with Assessor360, the overall performance of MTAOIQA* improves PLCC by $10.4\%$, SRCC by $10.8\%$, and RMSE by $13.1\%$ on the JUFE-10K database. Compared with other sub-optimal methods, MTAOIQA* has an increase of $3.9\%$, $9\%$, $5.3\%$ and $0.9\%$ in R1, R2, R3, and R4 in terms of PLCC on the OIQ-10K database. In addition, the performance of MTAOIQA* with auxiliary tasks has been further increased. It should be noted that MTAOIQA outperforms MTAOIQA* by $3.8\%$ and $6.2\%$ in terms of PLCC when facing the ST distortion and the R2 case, which are difficult to capture. 

\subsection{Cross-Database Evaluation}

We validate the generalizability of MTAOIQA by cross-database evaluation. Specifically, we train these models on one database and then test them on the other database. The experimental results are shown in Table~\ref{tab-cross-database}. From this table, we can clearly observe that the results of these models trained on the JUFE-10K database are worse than that on the OIQ-10K database. This is reasonable because the JUFE-10K database contains only non-uniform distortion, while the OIQ-10K database has both uniformly and non-uniformly distorted OIs. overall, the proposed MTAOIQA shows the best performance in this cross-database evaluation. 

\subsection{Ablation Studies}
\label{subsec:as}
In this section, we conduct a series of ablation experiments on the JUFE-10K and OIQ-10K databases to explore the effectiveness of the auxiliary tasks and these newly designed modules.

\paragraph{The effectiveness of auxiliary tasks} 

We choose different combinations of auxiliary tasks to train the proposed model, and the results on JUFE-10K and OIQ-10K databases are shown in Table~\ref{tab-choose-JUFE-10K} and \ref{tab-choose-OIQ-10K}. From these two tables, we have some interesting findings. First, the distortion range, type, and degree classification tasks all contribute to the quality prediction task. This indicates that the multitask learning model is more effective than the single-task model. Second, the performance does not necessarily improve with a higher number of auxiliary task selections. The results of selecting distortion range classification as an auxiliary task in Table~\ref{tab-choose-JUFE-10K} and selecting distortion range, type, and degree classification tasks as an auxiliary task are almost the same. This indicates that selecting auxiliary tasks corresponding to the features of different databases can improve the quality prediction. Furthermore, as can be clearly observed, the distortion range auxiliary task brings more performance gains than the distortion type and distortion degree auxiliary tasks, which indicates that the distribution of distortion can better express the characteristics of non-uniformly distorted OIs.

\begin{table}[H]
\caption{Cross database evaluation. The best performance results are highlighted in bold.}
\centering
\setlength{\tabcolsep}{3pt}
\begin{tabular}{c  c c c  c c c}
\hline
\toprule
\multirow{4}{*}{Models}  & \multicolumn{3}{c}{Train JUFE-10K}                           & \multicolumn{3}{c}{Train OIQ-10K} \\ & \multicolumn{3}{c}{Test OIQ-10K}                           & \multicolumn{3}{c}{Test JUFE-10K} \\ \cmidrule(lr){2-4} \cmidrule(lr){5-7} & \multicolumn{1}{c}{PLCC} & \multicolumn{1}{c}{SRCC} & RMSE & \multicolumn{1}{c}{PLCC} & \multicolumn{1}{c}{SRCC} & RMSE \\ 
\midrule 
MC360IQA~\cite{MC360IQA}                & 0.290 & 0.278 & 1.063 & 0.319 & 0.253 & 0.576\\ 
VGCN~\cite{VGCN}                        & 0.426 & 0.418 & 0.415 & 0.550 & 0.517 & 0.508\\ 
Fang22~\cite{fang2022perceptual}        & 0.274 & 0.162 & 0.441 & 0.429 & 0.366 & 0.549\\ 
Assessor360~\cite{wu2023assessor360}    & 0.357 & 0.367 & 0.428 & 0.624 & 0.614 & 0.475\\ 
MTAOIQA                                & \textbf{0.468} & \textbf{0.469} & \textbf{0.405} & \textbf{0.708} & \textbf{0.696} & \textbf{0.427}\\ 
\bottomrule
\end{tabular}
\label{tab-cross-database}
\end{table}

\begin{table}[H]
\caption{Experimental results of MTAOIQA with different auxiliary tasks or without any auxiliary tasks on the JUFE-10K database. R, T, and D denote the distortion range, distortion type, and distortion degree classification tasks, respectively. w/: with.}
\centering
\setlength{\tabcolsep}{3pt}
\begin{tabular}{c  c c c  c c c}
\hline
\toprule
\multirow{3}{*}{Task selection} & \multicolumn{6}{c}{JUFE-10K}  \\ \cmidrule(lr){2-7}  & \multicolumn{1}{c}{PLCC} & \multicolumn{1}{c}{SRCC} & RMSE & \multicolumn{1}{c}{ACC$_R$} & \multicolumn{1}{c}{ACC$_T$} & ACC$_D$ \\
\midrule 
w R+T+D                         & 0.819 & 0.818 & 0.347 & 0.999 & 0.957 & 0.909\\ 
w R+T                           & 0.803 & 0.803 & 0.360 & 0.992 & 0.954 & -\\ 
w R+D                           & \textbf{0.822} & \textbf{0.821} & \textbf{0.344} & 0.923 & - & 0.923\\ 
w T+D                           & 0.798 & 0.799 & 0.364 & - & 0.912 & 0.889\\ 
w R                             & 0.818 & 0.818 & 0.348 & 0.905 & - & -\\ 
w T                             & 0.807 & 0.810 & 0.357 & - & 0.998 & -\\ 
w D                             & 0.801 & 0.803 & 0.362 & - & - & 0.990\\
MTAOIQA*                              & 0.798 & 0.798 & 0.364 & - & - & -\\ 

\bottomrule
\end{tabular}
\label{tab-choose-JUFE-10K}
\end{table}

\begin{table}[H]
\caption{Experimental Results of MTAOIQA with or without the distortion range classification task on the OIQ-10K database. w/: with.}
\centering
\setlength{\tabcolsep}{3pt}
\begin{tabular}{c  c c c  c}
\hline
\toprule
\multirow{3}{*}{Task selection} & \multicolumn{4}{c}{OIQ-10K}  \\ \cmidrule(lr){2-5}  & \multicolumn{1}{c}{PLCC} & \multicolumn{1}{c}{SRCC} & RMSE & \multicolumn{1}{c}{ACC$_R$}\\
\midrule  
w R                             & \textbf{0.829} & \textbf{0.824} & \textbf{0.256} & 0.943 \\ 
MTAOIQA*                              & 0.819 & 0.813 & 0.263 & -\\ 

\bottomrule
\end{tabular}
\label{tab-choose-OIQ-10K}
\end{table}

\begin{table}[H]
\caption{Ablation study results of each component in the MTAOIQA* model on JUFE-10K and OIQ-10K databases. B: swin transform.}
\centering
\setlength{\tabcolsep}{3pt}
\begin{tabular}{c c c c  c c c  c c c}
\hline
\toprule
\multirow{2}{*}{B} & \multirow{2}{*}{MFS} & \multirow{2}{*}{MDI} & \multirow{2}{*}{MAF} & \multicolumn{3}{c}{JUFE-10K}        & \multicolumn{3}{c}{OIQ-10K} \\ \cmidrule(lr){5-7} \cmidrule(lr){8-10}
    &   &     &     & \multicolumn{1}{c}{PLCC} & \multicolumn{1}{c}{SRCC} & RMSE & \multicolumn{1}{c}{PLCC} & \multicolumn{1}{c}{SRCC} & RMSE \\
\midrule 
\checkmark &            &            &            & 0.451 & 0.407 & 0.539 & 0.727 & 0.725 & 0.315\\ 
\checkmark & \checkmark &            &            & 0.536 & 0.512 & 0.510 & 0.736 & 0.734 & 0.310\\ 
\checkmark &            & \checkmark &            & 0.502 & 0.456 & 0.522 & 0.716 & 0.707 & 0.320\\ 
\checkmark &            &            & \checkmark & 0.769 & 0.769 & 0.381 & 0.815 & 0.806 & 0.266\\ 
\checkmark & \checkmark & \checkmark &            & 0.778 & 0.782 & 0.379 & 0.771 & 0.772 & 0.291\\ 
\checkmark & \checkmark &            & \checkmark & 0.769 & 0.771 & 0.386 & 0.820 & 0.815 & 0.262\\ 
\checkmark &            & \checkmark & \checkmark & 0.763 & 0.765 & 0.390 & \textbf{0.828} & \textbf{0.822} & \textbf{0.257}\\ 
\checkmark & \checkmark & \checkmark & \checkmark & \textbf{0.798} & \textbf{0.798} & \textbf{0.364} & 0.819 & 0.813 & 0.263\\ 
\bottomrule
\end{tabular}
\label{tab-Effectiveness of model component}
\end{table}

\begin{table}[H]
\caption{Experimental results of MTAOIQA with different auxiliary tasks on the JUFE-10K database. R, T, and D denote the distortion range, distortion type, and distortion degree classification tasks, respectively. w/: with.}
\centering
\setlength{\tabcolsep}{2.5pt}
\begin{tabular}{c c c c c c c c c}
\hline
\toprule
\multirow{2}{*}{Task selection} & \multirow{2}{*}{MFS} & \multirow{2}{*}{MAF} & \multicolumn{6}{c}{JUFE-10K}  \\ \cmidrule(lr){4-9}  & & & \multicolumn{1}{c}{PLCC} & \multicolumn{1}{c}{SRCC} & RMSE & \multicolumn{1}{c}{ACC$_R$} & \multicolumn{1}{c}{ACC$_T$} & ACC$_D$ \\
\midrule 
w R+T+D                   & \checkmark &  & 0.714 & 0.712 & 0.423 & 0.996 & 0.921 & 0.896\\ 
    w R+T+D                   &  &  \checkmark & \textbf{0.825} & \textbf{0.825} & \textbf{0.342} & 0.993 & 0.941 & 0.919\\
w R+T+D                   & \checkmark &  \checkmark & 0.819 & 0.818 & 0.347 & 0.999 & 0.957 & 0.909\\
w R+T                     & \checkmark &  & 0.729 & 0.734 & 0.413 & 0.999 & 0.729 & -\\ 
w R+T                     &  &  \checkmark &0.818  & 0.821 & 0.348 & 0.993 & 0.817 & -\\
w R+T                   & \checkmark &  \checkmark & 0.803 & 0.803 & 0.360 & 0.992 & 0.954 & -\\
w R+D                   & \checkmark &  & 0.692 & 0.710 & 0.436 & 0.782 & - & 0.676\\ 
w R+D                   &  &  \checkmark & 0.812 & 0.811 & 0.352 & 0.845 & - & 0.908\\ 
w R+D                   & \checkmark &  \checkmark &0.822 & 0.821& 0.344 & 0.923 & - & 0.923\\
w T+D                   & \checkmark &  & 0.708 & 0.712 & 0.426 & - & 0.990 & 0.749\\ 
w T+D                   &  &  \checkmark & 0.809 & 0.809 & 0.355 & - & 0.910 &0.905 \\ 
w T+D                     & \checkmark & \checkmark & 0.798 & 0.799 & 0.364 & - & 0.912 & 0.889\\
w R                   & \checkmark &  & 0.726 & 0.727 & 0.415 & 0.830 & - &- \\ 
w R                 &  &  \checkmark & 0.822 & 0.822 & 0.344 & 0.952 & - & -\\ 
w R                       & \checkmark & \checkmark & 0.818 & 0.818 & 0.348 & 0.905 & - & -\\ 
w T                 & \checkmark &  & 0.703 & 0.699 & 0.430 & - & 1.000 & -\\ 
w T                   &  &  \checkmark & 0.808 & 0.806 & 0.356 & - & 0.994 & -\\ 
w T                       & \checkmark & \checkmark & 0.807 & 0.810 & 0.357 & - & 0.998 & -\\ 
w D                   & \checkmark &  & 0.754 & 0.759 & 0.397 & - & - & 0.811\\ 
w D                   &  &  \checkmark & 0.801 & 0.800 & 0.361 & - & - & 0.892\\ 
w D                       & \checkmark & \checkmark & 0.801 & 0.803 & 0.362 & - & - & 0.990\\

\bottomrule
\end{tabular}
\label{tab-JUFE-10K-MFS-MAF}
\end{table}

\begin{table}[H]
\caption{Experimental results of MTAOIQA with the distortion range auxiliary task on the OIQ-10K database. R denotes the distortion range classification task. w/: with.}
\centering
\setlength{\tabcolsep}{2.5pt}
\begin{tabular}{c c c c c c c c c}
\hline
\toprule
\multirow{2}{*}{Task selection} & \multirow{2}{*}{MFS} & \multirow{2}{*}{MAF} & \multicolumn{6}{c}{OIQ-10K}  \\ \cmidrule(lr){4-9}  & & & \multicolumn{1}{c}{PLCC} & \multicolumn{1}{c}{SRCC} & RMSE & \multicolumn{1}{c}{ACC$_R$} & \multicolumn{1}{c}{ACC$_T$} & ACC$_D$ \\
\midrule 
w R                   & \checkmark &  & 0.762  & 0.758 & 0.297 & 0.931& -& - \\ 
w R                   &  &  \checkmark & 0.828 & 0.823 & 0.257 & 0.926 & - & -\\ 
w R                  & \checkmark & \checkmark & \textbf{0.829} & \textbf{0.824} & \textbf{0.256} & 0.943 & - & -\\
\bottomrule
\end{tabular}
\label{tab-OIQ-10K-MFS-MAF}
\end{table}

\paragraph{Effectiveness of these specifically designed modules} 

To verify the effectiveness of these newly proposed MFS, MDI, and MAF modules, we conduct experiments on the JUFE-10K and OIQ-10K databases by removing one or some of them. The results are listed in Table~\ref{tab-Effectiveness of model component}, from which we can observe that the performance on the JUFE-10K and OIQ-10K databases improves with the addition of these individual modules. Among these modules, adding the MAF module alone obtains more gains on the two databases than other modules. This shows that the transformer can effectively capture both local and global features. The combination of the MFS, MDI, and MAF modules is beneficial in improving the performance of our proposed model, which indicates that the MFS module and MDI module can effectively capture non-uniform and uniform distortion and obtain better quality prediction results. In addition, we conduct experiments to further verify the effect of the MFS and MAF modules on different auxiliary tasks, and the experimental results are shown in Tables~\ref{tab-JUFE-10K-MFS-MAF} and \ref{tab-OIQ-10K-MFS-MAF}. From these Tables, we can observe that the MAF module greatly affects the performance of MTAOIQA, which indicates that the performance of MTAOIQA largely depends on the interaction between the auxiliary task and the main task.

\begin{table}[H]
\caption{The Performance of MTAOIQA* With different viewport generation methods.}
\centering
\setlength{\tabcolsep}{2pt}
\begin{tabular}{c  c c c c  c c c}
\hline
\toprule
\multirow{3}{*}{Generate viewports} & \multirow{3}{*}{Num} & \multicolumn{3}{c}{JUFE-10K}  & \multicolumn{3}{c}{OIQ-10K} \\ \cmidrule(lr){3-5} \cmidrule(lr){6-8} & &\multicolumn{1}{c}{PLCC} & \multicolumn{1}{c}{SRCC} & RMSE & \multicolumn{1}{c}{PLCC} & \multicolumn{1}{c}{SRCC} & RMSE \\ 
\midrule 
Spherical sampling~\cite{fang2022perceptual}               & 20 & 0.815 & 0.816 & 0.349 & 0.820 & 0.814 & 0.263\\ 
Spherical sampling~\cite{fang2022perceptual}               & 8  & 0.768 & 0.759 & 0.387 & 0.800 & 0.794 & 0.275\\ 
Saliency~\cite{VGCN}                                       & 20 & 0.812 & 0.812 & 0.3521 & 0.803 & 0.796 & 0.274\\ 
Saliency~\cite{VGCN}                                       & 8  & 0.712 & 0.699 &0.424 & 0.788 & 0.777 & 0.282\\ 
RPS~\cite{wu2023assessor360}                               & 20 & 0.385 & 0.270 & 0.557 & 0.616 & 0.568 & 0.361\\ 
RPS~\cite{wu2023assessor360}                               & 8  & 0.276 & 0.224 & 0.580 & 0.500 & 0.427 & 0.400\\
Equatorial sampling                                        & 20 & \textbf{0.817} & \textbf{0.817} & \textbf{0.348} & \textbf{0.831} & \textbf{0.829}& \textbf{0.255} \\ 
Equatorial sampling                                       &8  & 0.798 & 0.798 & 0.364 & 0.819 & 0.813 & 0.263\\ 

\bottomrule
\end{tabular}
\label{tab-viewport generation methods}
\end{table}

\paragraph{Influence of viewport generation methods} 
We explore the influence of different viewport sampling methods as well as the number of extracted viewports. As shown in Table~\ref{tab-viewport generation methods}, we can clearly observe that RPS~\cite{wu2023assessor360} performs poorly compared to other methods. It may be that the RPS~\cite{wu2023assessor360} method samples viewports in a limited region and would ignores those distorted regions. In addition, the equatorial sampling method achieves better results. This indicates that the information in the low-latitude region of OI can effectively represent the overall information. Besides, we can find that increasing the number of viewports helps MTAOIQA* to improve its performance, which is attributed to that more viewports bring richer information to MTAOIQA*, and therefore lead to more accurate quality predictions. 

\section{Conclusion}
\label{sec:conc}

In this paper, we propose a multitask auxiliary OIQA model named MTAOIQA for non-uniformly distorted omnidirectional images, which uses the swin transformer to extract multi-scale features of viewports, and designs a MFS module for dynamically allocating specific features to different tasks with learnable probabilities, a series of optional attention modules for enhancing perspective representation ability and a MAF module for aggregating those quality-aware features from different branches. We compare the proposed MTAOIQA with 16 state-of-the-art quality metrics on the JUFE-10K and OIQ-10K databases, and extensive experiments demonstrate that MTAOIQA achieves excellent performance in capturing non-uniform distortion, and benefits from these auxiliary tasks.

%



\ifCLASSOPTIONcaptionsoff
  \newpage
\fi



%

\bibliographystyle{IEEEtran}
\bibliography{reference}

\begin{IEEEbiography}[{\includegraphics[width=1.0in,height=1.3in,clip]{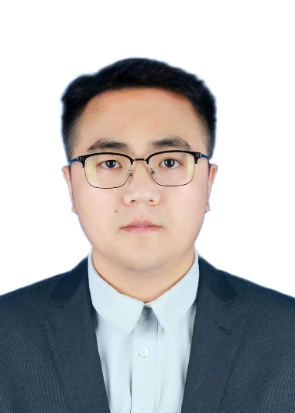}}]{Jiebin Yan} received the Ph.D. degree from Jiangxi University of Finance and Economics, Nanchang, China. He was a computer vision engineer with MTlab, Meitu. Inc, and a research intern with MOKU Laboratory, Alibaba Group. From 2021 to 2022, he was a visiting Ph.D. student with the Department of Electrical and Computer Engineering, University of Waterloo, Canada. He is currently a Lecturer with the School of Computing and Artificial Intelligence, Jiangxi University of Finance and Economics, Nanchang, China. His research interests include visual quality assessment and computer vision.
\end{IEEEbiography}

\vspace{-10 mm} 

\begin{IEEEbiography}[{\includegraphics[width=1.0in,height=1.3in,clip]{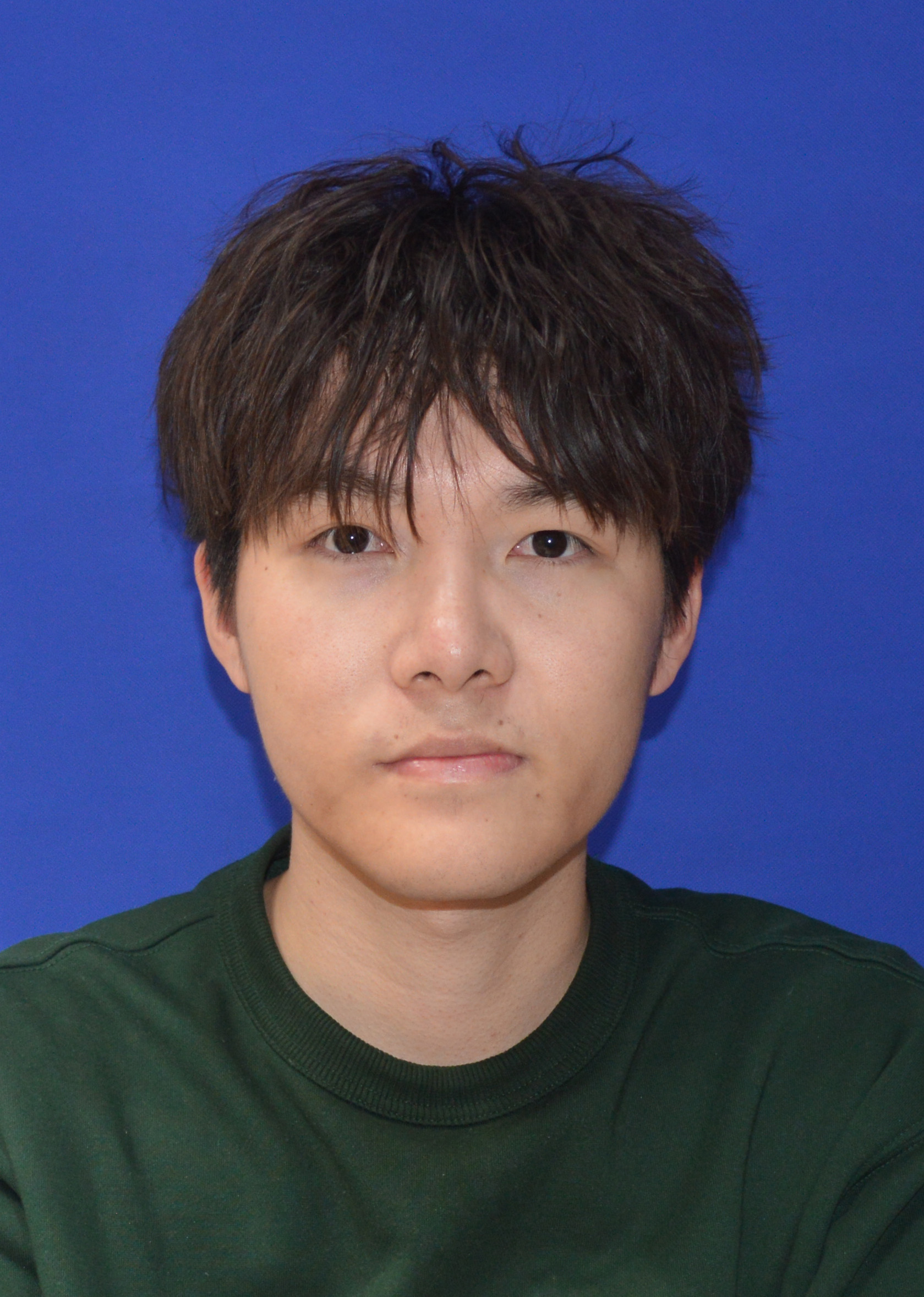}}]{Jiale Rao} received the B.E. degree from the Jiangxi Agricultural University, Nanchang, China, in 2022,  He is currently working toward the M.S. degree with the School of Computing and Artificial Intelligence, Jiangxi University of Finance and Economics. His research interests include visual quality assessment and VR image processing.
\end{IEEEbiography}

\vspace{-10 mm} 

\begin{IEEEbiography}[{\includegraphics[width=1.0in,height=1.3in,clip]{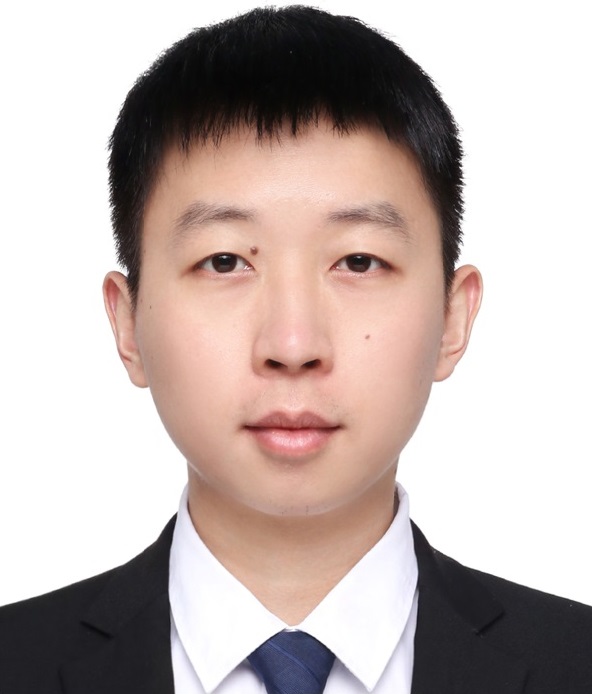}}]{Junjie Chen} received the B.E. degree in 2018 from Sichuan University, Sichuan, China, and Ph.D degree in 2023 from Shanghai Jiao Tong University, Shanghai, China. He is currently a Lecturer with the School of Computing and Artificial Intelligence, Jiangxi University of Finance and Economics, Nanchang, China. His current research interests include computer vision, deep learning, and machine learning.
\end{IEEEbiography}

\vspace{-10 mm} 

\begin{IEEEbiography}[{\includegraphics[width=1.0in,height=1.3in,clip]{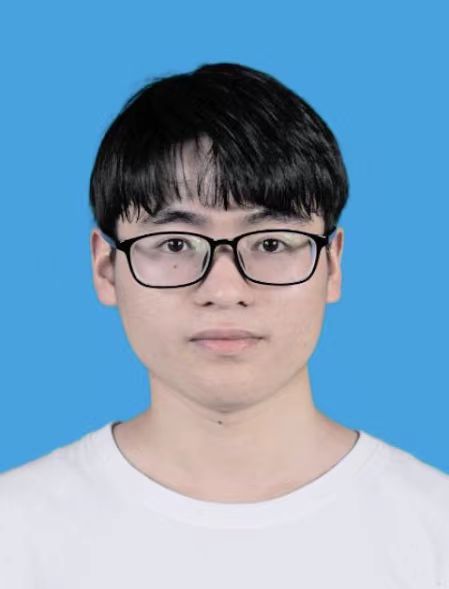}}]{Ziwen Tan} received the B.E. degree from the School of Modern Economics and Management, Jiangxi University of Finance and Economics, Nanchang, China, in 2022. He is currently working toward the M.S. degree with the School of Computing and Artificial Intelligence, Jiangxi University of Finance and Economics. His
research interests include visual quality assessment and VR image processing.
\end{IEEEbiography}
\vspace{-10 mm} 

\begin{IEEEbiography}[{\includegraphics[width=1.0in,height=1.3in,clip]{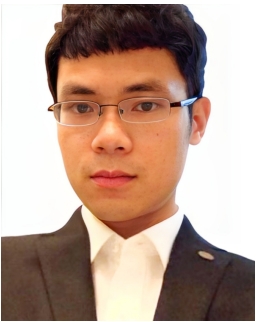}}]{Weida Liu} received the B.S. degree in computer science from Nanyang Technological University, Singapore. He is currently pursuing the Ph.D. degree with School of Computer Science and Engineering, Nanyang Technological University, Singapore. And he is also a senior research engineer with the Institute for Infocomm Research, Agency for Science, Technology and Research, Singapore. His research interests include image segmentation and computer vision.
\end{IEEEbiography}

\vspace{-10 mm} 

\begin{IEEEbiography}[{\includegraphics[width=1.0in,height=1.3in,clip]{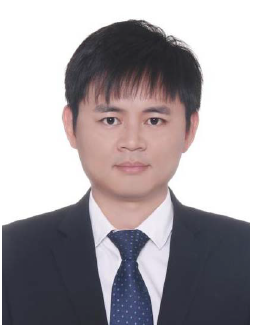}}]{Yuming Fang}(S’13–SM’17) received the B.E. degree from Sichuan University, Chengdu, China, the M.S. degree from the Beijing University of Technology, Beijing, China, and the Ph.D. degree from Nanyang Technological University, Singapore. He is currently a Professor with the School of Computing and Artificial Intelligence, Jiangxi University of Finance and Economics, Nanchang, China. His research interests include visual attention modeling, visual quality assessment, computer vision, and 3D image/video processing. He serves on the editorial board for \textsc{IEEE Transactions on Multimedia} and \textsc{Signal Processing: Image Communication}.
\end{IEEEbiography}

\end{document}